\documentclass[aps,prb,showpacs,amsfonts,amssymb,amsmath]{revtex4}

\usepackage{amsfonts}
\usepackage{amsmath}
\usepackage{amssymb}
\usepackage{graphicx}
\begin{document}

\title{Persistent spin current in nano-devices and definition of the spin current}

\author{Qing-feng Sun$^{\ast}$\\
{\sl\small Beijing National Lab for Condensed Matter Physics and
Institute of Physics, Chinese Academy of Sciences, Beijing 100080, China \\
}
{X. C. Xie}\\
{\sl\small Department of Physics, Oklahoma State University,
Stillwater, Oklahoma 74078;\\
Beijing National Lab for Condensed Matter Physics and Institute of
Physics, Chinese Academy of Sciences, Beijing 100080,
China } \\
{ Jian Wang}\\
{\sl\small Department of Physics and the center of theoretical and
computational physics, The University of Hong Kong, Hong Kong, China
 } }

\date{\today}

\begin{abstract}

We investigate two closely related subjects: (i) existence of a pure
persistent spin current in semiconducting mesoscopic device with a
spin-orbit interaction's (SOI), and (ii) the definition of the spin
current in the presence of SOI. Through physical argument from four
physical pictures in different aspects, we provide strong evidences
that the persistent spin current does exist in a device with SOI in
the absence of any magnetic materials. This persistent spin current
is an analog of the persistent charge current in a mesoscopic ring
threaded by a magnetic flux, and it describes the real spin motion
and can be measured experimentally. We then investigate the
definition of the spin current. We point out that (i) the non-zero
spin current in the equilibrium SOI device is the persistent spin
current, (ii) the spin current is in general not conserved, and
(iii) the Onsager relation is violated for the spin transport no
matter what definition of the spin current is used. These issues,
the non-zero spin current in the equilibrium case, the non-conserved
spin current, and the violation of the Onsager relation, are
intrinsic properties of spin transport. We note that the
conventional definition of the spin current has very clear physical
intuition and describes the spin motion very well. Therefore we feel
that the conventional definition of the spin current makes physical
sense and there is no need to modify it. In addition, the
relationship between the persistent spin current and transport spin
current, the persistent linear and angular spin currents in the SOI
region of the hybrid ring, are discussed. Finally, we show that if
the spin-spin interaction is included into the Hamiltonian, the
persistent spin current is automatically conserved using the
conventional definition.

\end{abstract}


\maketitle

\section{Introduction}

In traditional charge-based electronics, the spin degree of freedom
has not been fully explored. Until recently, the scientists found
that spin control and manipulation in the nano-scale can enhance
operational speed and integration density of conventional
charge-based electronic devices.\cite{aref2,aref3} In order to
successfully use the spin degree of freedom of electrons in the
conventional semiconductor devices, one has to have a good control,
manipulation, and detection of the spin and its flow in
nano-devices. This emerging field called spintronics is a new
sub-discipline of condensed matter physics, and it is growing
rapidly and generating great interests in recent
years.\cite{aref2,aref3}

It is well known that the spin-orbit interaction (SOI) plays an
important role in the emerging field of semiconductor spintronics.
SOI couples the spin degree of freedom of electrons to their orbital
motions, thereby giving rise to a useful way to manipulate and
control the spin of electron by an external electric field or a gate
voltage. SOI is an intrinsic interaction having its origin from
relativistic effects that can be quite significant in some
semiconductors.\cite{addref1} For instance experiments show that the
significant SOI indeed exist in some
semiconductors,\cite{aref4,aref5,aref6,aref7,aref8} e.g. Rashba SOI
was found in the InGaAs/InAlAs heterostructure or InAs Quantum
Wells,\cite{aref4,aref5} and both the Rashba and Dresselhaus SOI
were detected in various III-V material based 2D structures at room
temperature,\cite{aref6} to just name a few. Moreover, the strengths
of these SOI have been well modulated by the gate voltage
experimentally.\cite{aref7}

Many interesting effects resulting from SOI have been predicted. For
example, using the effect of spin precessions due to the Rashba SOI,
Datta and Das proposed a spin-transistor more than ten years
ago.\cite{aref9} By using the Rashba SOI, a method to efficiently
control and manipulate the spin of the electron in the quantum dot
has been proposed.\cite{aref10} Very recently, a very interesting
effect, the intrinsic spin Hall effect, is theoretically predicted
by Murakami {\it et.al.} and Sinova {\sl et.al.} in a Luttinger SOI
3D p-doped semiconductor and a Rashba SOI 2D electron
gas,\cite{aref11,aref12} respectively, that a substantial amount of
dissipationless spin current can be generated from the interplay
between the electric field and the SOI. Since then, the spin Hall
effect has generated tremendous interests with a great amount of
works focusing in the field of
spintronics.\cite{aref13,aref14,aref15,aref16,aref17,aref18} On
experimental side, two groups by Kato {\sl et al.}\cite{aref19} and
Wunderlich {\sl et al.}\cite{aref20} have observed the transverse
opposite spin accumulations near two edges of their devices when the
longitudinal voltage bias is added. In additional, a third group by
Valenzuela and Tinkham took the electric measurement of the
reciprocal spin Hall effect,\cite{aref21} and they have observed an
induced transverse voltage in a diffusive metallic conductor when a
longitudinal net spin current flows through it.

In this paper, we study two closely related subjects that: (i) we
predict another interesting effect that a persistent spin current
without accompanying charge current exists in a coherent
mesoscopic semiconductor ring with symplectic symmetry, i.e., with
SOI but maintaining the time reversal symmetry; (ii) we examine
the issue whether it is needed to redefine the conventional spin
current, or in other words, whether the conventional definition of
the spin current, ${\bf I}_S = Re \{ \Psi^{\dagger} \hat{v}
\hat{s} \Psi \}$,\cite{note2} is reasonable in the presence of
SOI. Since these two subjects are closely related, we investigate
them together here. In fact, a few years ago, by using the
conventional definition of the spin current Rashba has found a
nonzero spin current $I_S$ in an infinite two-dimensional system
with Rashba SOI in the thermodynamic equilibrium.\cite{aref23} In
his opinion, this spin current $I_S$ is not associated with real
spin transport, and therefore should be eliminated in calculating
the transport current by modifying the conventional definition of
the spin current. After his work, many subsequent work have
discussed the definition of the spin
current.\cite{aref24,aref25,aref26,aref27} So to uncover the
physical meaning of the non-zero spin current in the equilibrium
system, we must face the question of whether one needs to redefine
the spin current.

More than two decades ago, the persistent (charge) current in a
mesoscopic ring threaded by a magnetic flux has been predicted
theoretically,\cite{aref28} and later observed experimentally in
early 1990s.\cite{aref29} It is now well known that the persistent
charge current is a pure quantum effect and can sustain without
dissipation in the equilibrium case. There has also been many
investigations on the persistent spin
current.\cite{aref30,aref31,aref32,aref33,aref34} For example, in a
mesoscopic ring with a crown-shape inhomogeneous magnetic
field\cite{aref30} or threaded by a magnetic flux\cite{aref31}, the
persistent spin current has been predicted and is related to the
Berry's phase. Recently, the persistent spin current carried by
Bosonic excitations has also been predicted in a Heisenberg ring
with the magnetic field or in the ferromagnetic
material.\cite{aref32} The reason that the persistent spin current
exists may be explained as follows. Due to the magnetic field or the
magnetic flux, there are persistent flows of both spin up and down
electrons. In the absence of SOI, this gives rise to the well known
persistent charge current. In the presence of SOI or magnetic field,
the persistent charge current is spin polarized resulting a nonzero
persistent spin current. Hence the origin of this persistent spin
current is the same as that of persistent charge current so that the
persistent spin current always accompanies with a persistent charge
current. Recently,\cite{aref35,aref39} we have reported that a
persistent spin current without accompanying charge current (a pure
persistent spin current) can sustains in a mesoscopic semiconducting
ring with SOI in the absence of the external magnetic field,
magnetic flux, and the magnetic material. This pure persistent spin
current is induced solely by SOI, which is different with the
pervious case. In main part of this paper, we will give detailed
discussions on the existence of the persistent spin current and the
behavior of the spin current.

Another motivation of the present paper is to study the definition
of the spin current in the presence of SOI. In fact, the spin
current is the most important physical quantity in the field of the
spintronics. So it is important to give an appropriate definition of
the spin current. This problem, the discussion of the definition of
the spin current, is first tackled by Rashba. In a recent work by
Rashba,\cite{aref23} he found a nonzero spin current $I_S$ in an
infinite two-dimensional system with Rashba SOI even in the
equilibrium case by using the conventional definition of the spin
current,\cite{note2} ${\bf I}_S = Re \Psi^{\dagger} \hat{v} \hat{s}
\Psi$. So he questioned the correctness of the conventional
definition of the spin current and suggested that the conventional
definition should be modified to eliminate the equilibrium non-zero
spin current. Besides the nonzero spin current in the equilibrium,
there are others problems with the conventional definition: (1) the
spin current is not conserved using the conventional definition in
the presence of SOI or/and the magnetic field. This is because the
operator $\hat{s}$ does not commute with the Hamiltonian $H$ when
the system has SOI or/and the magnetic field. (2). the Onsager
relation is violated using the conventional definition. After
Rashba's work, many subsequent papers have discussed the definition
of the spin current.\cite{aref24,aref25,aref26,aref27} For instance,
Sun and Xie suggested that in addition to the conventional (linear)
spin current ${\bf I}_S = Re \Psi^{\dagger} \hat{v} \hat{s} \Psi$,
one needs to introduce the angular spin current ${\bf I}_{\omega} =
Re\Psi^{\dagger} (d\vec{s}/dt)\Psi = Re \Psi^{\dagger} \hat{\omega}
\times \hat{s} \Psi$ to describe the rotational motion (precession)
of a spin,\cite{aref24} because the spin operator is a vector.
Similar to the conventional linear spin current, the angular spin
current can also induce an electric field. Shi {\sl et.
al.}\cite{aref26} gave a new definition of the spin current with
${\bf I}_S = Re \Psi^{\dagger} d( {\bf r} \hat{s})/dt \Psi$, in
which the operator of the spin current $\hat{I}_S = d( {\bf r}
\hat{s})/dt$ is the derivative of the whole ${\bf r} \hat{s}$.
Compared with the conventional definition of the spin current, it
has an extra term ${\bf r} (d\hat{s}/dt)$ in this new definition.
Wang {\sl et. al.}\cite{aref27} pointed out that the spin current is
automatically conserved using the conventional definition of the
spin current if the spin-spin interaction is included. In the Sec.
VI in the present paper, we will explore the definition of the spin
current.

In this paper, we first give physical argument as to why the
persistent spin current can exist in equilibrium. In order to show
that the persistent spin current should exist in the mesoscopic
system with solely SOI, four physical pictures or physical argument
from the different aspects will be discussed: (i) from the picture
of electric and magnetic correspondence to analyze the driving force
of persistent spin current (in Fig.1a and Fig.1b); (ii) from the
point of view of the spin Berry phase; (iii) from the comparison
among four effects, the Hall effect, spin Hall effect, the
persistent (charge) current, and the persistent spin current (see
Fig.2); (iv) from the point of view of the motion of the spin. As an
example, we then consider a semiconducting SOI-normal hybrid
mesoscopic ring and show that indeed a pure persistent spin current
can emerge with solely SOI. We note that currently there is no
consensus on the definition for the spin current in the presence of
SOI. In order to avoid the problem of the definition of the spin
current, here we first use the following approach: we consider a
mesoscopic hybrid ring that consists of a Rashba SOI's region and a
normal region without SOI as shown in Fig.1c. Since there is no
spin-flip in this normal region, the definition of the spin current
in that region is without controversy. So we can calculate and study
the spin current in the normal region, and to make sure that the
persistent spin current indeed can be induced solely by SOI. After
making sure the existence of the persistent spin current, we then to
investigate the definition of the spin current. (i). We point out
that the non-zero spin current in the equilibrium system in the
presence of the SOI is the persistent spin current. It describes a
real spin motion and has the physical meaning, so this spin current
should be kept as it is. (ii). Since spin operator is a vector, it
has the rotational motion (precession) due to the SOI in accompany
with the particle translational motion, so in general spin current
is not a conserved quantity. In fact, there are experimental
indications that the spin current is not conserved.\cite{aref21} On
the other hand, in certain cases if one includes the strong
spin-spin interaction into the Hamiltonian, the spin current
obtained from the conventional definition will be conserved
automatically. (iii). It is well known that the Onsager relation
holds under certain condition. We will show that for the spin
system, this condition does not satisfy. We think that the above
three points are actually the intrinsic properties of spin transport
so we feel that the conventional definition of the spin current need
not to be modified.\cite{note1}

In addition, we also address the following issues. (1). The relation
between the persistent spin current and transport spin current. We
note that they can not be distinguished from each other in the
coherent part of the device. (2). In calculating the persistent
linear and angular spin currents in the SOI's region of the
SOI-normal hybrid ring, we find that the persistent spin current
still exists in the SOI's region even if the whole ring has the SOI.
So the normal part is not necessary for generating the persistent
spin current. (3). The measurement of the persistent spin current is
discussed, we suggest that this persistent spin current can be
observed by detecting its induced electric fields. (4). Including
the spin-spin interaction in the ring, we find that the persistent
spin current calculated using the conventional definition is a
conserved quantity.

The rest of the paper is organized as follows. In Section II, we
provide physical arguments and physical pictures from four different
aspects to show the existence of the persistent spin current. In
Section III, we consider a SOI-normal hybrid mesoscopic
semiconducting ring device to show that indeed a pure persistent
spin current can emerge in the normal region where the definition of
the spin current is without controversy. Then the effect of sharp
interface between the SOI's region and the normal region, and the
relation of the persistent spin current and the transport spin
current, are discussed in Section IV and V, respectively. In Section
VI, we study the definition of the spin current. In Section VII, the
persistent linear and angular spin current in the SOI's region as
well some conserved quantities in the hybrid ring device are
explored. In Section VIII, we investigate electric fields induced by
the persistent spin current. This provides a way to detect the
persistent spin current. In Section IX, we discuss the effect of
spin-spin interaction in the ring which makes the persistent spin
current conserved. Finally, Section X summarizes the results of our
work.

\section{Physical arguments for the existence of the
persistent spin current}
\label{sec2}

In this section, we argue that the pure persistent spin current
(without accompanying charge current) should exist in the mesoscopic
semiconducting ring device with the SOI. In particularly, this pure
persistent spin current can be induced solely by SOI even at zero
magnetic flux or magnetic field. We examine this new effect from
following four different aspects.

\subsection{Analysis of the driving force}

For the persistent charge current in the mesoscopic ring, the
magnetic flux or magnetic field acts like a "driving force",  so one
naturally looks for the analogous "driving force" in the spin case.
To discuss this question, let us consider two devices. The first
device consists of a mesoscopic ring (without SOI) where a magnetic
atom with a magnetic dipole moment is placed at the center of the
ring (see Fig.1a). In the second device the magnetic atom is
replaced by a charged atom, e.g., an ion (see Fig.1b). The magnetic
atom produces a vector potential ${\bf A}$ on the perimeter of the
ring which drives the persistent charge current. By analogy, a
charged atom which produces a scalar potential $\phi$ on the
perimeter of the same ring should drive a persistent spin
current.\cite{note3} Since the presence of this ionic center
generates a SOI in the relativistic limit, we expect that this SOI
which plays the role of the spin "driving force" will induce a pure
persistent spin current. In addition, from the picture of electric
and magnetic correspondence, the persistent charge current in the
ring should also change into the persistent spin current, when the
center magnetic atom (in Fig.1a) is substituted by a charged atom
(in Fig.1b).

\subsection{From the point of view of the spin Berry phase}

Second, the existence of the pure persistent spin current can be
examined from another point of view using the spin Berry
phase.\cite{aref37,aref38} This physical arguments has been provided
in detail in our previous Letter (see the 4-th paragraph in
Ref.\cite{aref35}), so we omit the discussion here.

% It is well known that an electron circulating a ring with a
% non-uniform magnetic field or magnetic flux acquires a geometric
% phase the so-called Berry phase.\cite{aref37} It has been discovered
% by Loss et al that it is this Berry phase $\chi$ that induces the
% well known persistent charge current.\cite{aref30} Assuming that the
% electron wavelength is much smaller than the perimeter of the ring
% and the electron motion is quasi-classical, let us examine an
% electron with spin $\sigma$ traverses slowly along the ring with
% only SOI.\cite{aref38} Due to the SOI, the spin of this electron
% precesses and acquires a geometric phase after the electron returns
% to its starting point.\cite{aref38} This is the so-called spin Berry
% phase\cite{aref38}. The spin Berry phase due to Rashba SOI for an
% electron with spin $\sigma$ moving in the clockwise direction is
% found to be\cite{aref38} $\chi_\sigma = \sigma\pi$ where $\sigma =
% \pm$ for $\sigma = \uparrow,\downarrow$. From the physical picture
% due to Loss et al,\cite{aref30} the spin Berry phase $\chi_+$ for
% the spin up electron induces a clockwise persistent spin polarized
% current $I_1$. Similarly, the Berry phase $\chi_-$ induces a
% counter-clockwise persistent spin polarized current with the
% polarization exactly opposite to that of $I_1$ since our system has
% time reversal symmetry. As a result, the spin Berry phase due to SOI
% will induce a pure persistent spin current.

\subsection{Comparison among four effects: the Hall effect, spin Hall effect, the
persistent (charge) current, and the persistent spin current}

In the following, let us compare the four effects: the Hall effect,
the spin Hall effect, the persistent charge current, and the
persistent spin current, from which one expect that the persistent
spin current should exist in the mesoscopic ring with solely SOI.

(i) We consider a four-terminal device with a bias $V$ added between
the terminals 1 and 3 and a perpendicular magnetic field ${\bf B}$
(as shown in Fig.2a). For this system there exists the Hall effect,
and a charge current is induced in the transverse terminals 2 and 4.
(ii) Replacing the perpendicular magnetic field by the SOI (see
Fig.2b), a spin current emerges (instead of the charge current) in
the transversal terminals 2 and 4. This is the spin Hall effect
predicted recently and generated tremendous
interests.\cite{aref11,aref12,aref13,aref14,aref15,aref16,aref17}
(iii) Consider a mesoscopic ring with a perpendicular magnetic field
(see Fig.2c), a persistent charge current is induced in the ring.
This is the persistent current, which is well known
now.\cite{aref28,aref29} (iv) By analogy, a persistent spin current
instead of the persistent charge current should be induced when the
SOI replaces the perpendicular magnetic field (see Fig.2d).

Let us discuss the Hamiltonian in four devices in Fig.2. (i) In
Fig.2a, the Hamiltonian is $H=\frac{({\bf p} + e {\bf A}/c)^2}{2m} +
V({\bf r})$, and there exists the Hall effect because of the vector
potential ${\bf A}$. (ii) To replace ${\bf A}$ by the Rashba SOI
with $H=\frac{({\bf p} + \alpha \hat{\sigma}\times \hat{z})^2}{2m} +
V({\bf r})$, the Hall effect change into the spin Hall effect, and a
spin current instead of the charge current emerges in the terminals
2 and 4 (see Fig.2b). (iii) In Fig.2c, the Hamiltonian is
$H=\frac{({\bf p} + e {\bf A}/c)^2}{2m} + V({\bf r})$, and there
exist the persistent charge current in the ring because of the
vector potential ${\bf A}$. (iv) By analogy, when ${\bf A}$ is
replaced by the Rashba SOI, a persistent spin current should emerge
instead of the persistent charge current.

\subsection{From point of view of the motion of the spin}

By analyzing the motion of the spin, including the translational and
rotational motion (precession) of a spin, one can also show the
existence of the persistent spin current in the mesoscopic device
with the SOI.\cite{aref39} In order to analyze the motion of the
spin, we need to solve the wave-functions of the electron, so we put
this analysis in the appendix, where the wave-functions have been
solved.

\section{Persistent spin current in the normal region of the SOI-normal hybrid ring}
\label{sec3}

In this section, we present an example to show that indeed a pure
persistent spin current can exist for a mesoscopic semiconducting
ring with SOI. In the presence of SOI, the spin of an electron
experiences a torque and hence $\sigma_i$ ($i=x,y,z$) is not a good
quantum number anymore. Because of this, the spin current is not
conserved using the conventional definition. At present there are
controversies on whether one should define a conserved spin current
or whether there exists a conserved spin
current.\cite{aref24,aref25,aref26,aref27} In another word, so far
there is no consensus on the definition for the spin current in the
presence of SOI. In this section, we use the following approach. We
first discuss the persistent spin current for a one-dimensional
mesoscopic semiconducting ring that consists of a Rashba SOI's
region and a normal region without SOI as shown in Fig.1c. Since
there is no spin-flip in the normal region, the spin current can be
calculated using conventional definition without controversy, so
that we can make sure the existence of the persistent spin current
in the equilibrium case with solely SOI. After we make sure the
existence of the persistent spin current, we then go back to examine
the definition of the spin current, which is given in Section VI.

The Hamiltonian of our system is given by:\cite{aref31,aref40}
\begin{equation}
  H = -E_a \frac{\partial^2}{\partial \varphi^2}
    -\frac{i\sigma_{r}}{2a}\left[
  \alpha_R(\varphi) \frac{\partial}{\partial \varphi}
 + \frac{\partial}{\partial \varphi} \alpha_R(\varphi)
 \right]
  -i \frac{\alpha_R(\varphi)}{2a}\sigma_{\varphi}
\end{equation}
where $E_a =\hbar^2/2ma^2$, $a$ is the radius of the ring, $m$ is
the effective mass of the electron, $\sigma_r = \sigma_x \cos\varphi
+\sigma_y \sin\varphi$, and $\sigma_{\varphi} = -\sigma_x
\sin\varphi +\sigma_y \cos\varphi$. $\alpha_R(\varphi)$ is the
strength of the Rashba SOI, $\alpha_R(\varphi)=0$ while $0 <\varphi
<\Phi_0$, i.e. in the normal region, and $\alpha_R(\varphi)$ is a
constant $\alpha_R$ in the SOI's region with $\Phi_0 <\varphi
<2\pi$.

The eigenstates of Hamiltonian (1) can be solved numerically in the
following way.  First in the Rashba SOI's region ($\alpha_R \not=
0$), the equation $H\Psi(\varphi)=E \Psi(\varphi)$ has four
independent solutions $\Psi^{SO}_{i}(\varphi)$
($i=1,2,3,4$):\cite{aref31}
\begin{equation}
 \Psi^{SO}_{1/2}(\varphi) = \left(\begin{array}{l}
       \cos (\theta/2) e^{ik_{1/2}\varphi} \\
       -\sin (\theta/2) e^{i(k_{1/2}+1)\varphi}
       \end{array}
       \right),
\end{equation}
and $\Psi^{SO}_{3/4} = \hat{T} \Psi^{SO}_{1/2}$ with $\hat{T}$ being
the time-reversal operator. In Eq.(2), the wave vectors $k_{1/2}=
-1/2 +1/(2\cos\theta) \pm (1/2) \sqrt{ (1/\cos^2\theta)-1 +4E/E_a}$,
and the angle $\theta$ is given by $\tan(\theta)= \alpha_R/(a E_a)$.
Similarly, in the normal region ($0 <\varphi <\Phi_0$), the
Sch$\ddot{o}$dinger equation has four independent solutions:
$\Psi^N_{1}(\varphi) = (1,0)^{\dagger} e^{ik\varphi}$,
$\Psi^N_{2}(\varphi) = (1,0)^{\dagger} e^{-ik\varphi}$, and
$\Psi^{N}_{3/4} = \hat{T} \Psi^{N}_{1/2}$, with $k=\sqrt{E/E_a}$.
Secondly, the eigen wave function $\Psi(\varphi)$ with the eigen
energy $E$ can be represented as:
\begin{eqnarray}
  \Psi(\varphi) = \left\{ \begin{array}{l}
            \sum\limits_{i} a_i \Psi^{N}_i(\varphi) ,   \hspace{6mm}
             while \hspace{2mm} 0 < \varphi < \Phi_0 \\
            \sum\limits_{i} b_i \Psi^{SO}_i(\varphi) ,   \hspace{5mm}
             while \hspace{2mm} \Phi_0 < \varphi < 2\pi
             \end{array} \right.
\end{eqnarray}
where $a_i$ and $b_i$ ($i=1,2,3,4$) are constants to be determined
by the boundary conditions at the interfaces $\varphi=0$ and
$\Phi_0$.  Here the boundary conditions are the continuity of the
wave function $\Psi(\varphi)|_{\varphi=0^+/\Phi_0^+} =
\Psi(\varphi)|_{\varphi=2\pi^-/\Phi_0^-} $ and the continuity of its
flux\cite{note4} $\hat{v}_{\varphi} \Psi|_{\varphi=0^+/\Phi_0^+} =
 \hat{v}_{\varphi} \Psi|_{\varphi=2\pi^-/\Phi_0^-} $, where
$\hat{v}_{\varphi} =(2aE_a/i\hbar) [\partial/\partial \varphi +(i/2)
\sigma_r \tan(\theta)] $ is the velocity operator. By using the
boundary conditions, we obtain eight series of linear equations:
\begin{eqnarray}
&& a_1+a_2 + \cos (\theta/2) e^{i2\pi k_1} b_1
 +\cos (\theta/2) e^{i2\pi k_2} b_2 \nonumber \\
 &&
 +\sin (\theta/2) e^{-i2\pi k_1} b_3
 +\sin (\theta/2) e^{-i2\pi k_2} b_4 =0
 \end{eqnarray}
 \begin{eqnarray}
&& a_3+a_4 - \sin (\theta/2) e^{i 2\pi k_1} b_1
  -\sin (\theta/2) e^{i 2\pi k_2} b_2 \nonumber \\
 &&+\cos (\theta/2) e^{-i2\pi k_1} b_3
 +\cos (\theta/2) e^{-i2\pi k_2} b_4 =0
 \end{eqnarray}
 \begin{eqnarray}
&& e^{ik\Phi_0}a_1 + e^{-ik\Phi_0}a_2
 + \cos (\theta/2) e^{i k_1 \Phi_0} b_1
 +\cos (\theta/2) e^{i k_2 \Phi_0} b_2 \nonumber \\
 &&+\sin (\theta/2) e^{-i (k_1 +1)\Phi_0} b_3
 +\sin (\theta/2) e^{-i (k_2 +1)\Phi_0} b_4 =0
 \end{eqnarray}
 \begin{eqnarray}
&& e^{-ik\Phi_0} a_3 + e^{ik\Phi_0}a_4
 - \sin (\theta/2) e^{i (k_1 +1)\Phi_0} b_1
 - \sin (\theta/2) e^{i (k_2 +1)\Phi_0} b_2 \nonumber \\
 &&+\cos (\theta/2) e^{-i k_1 \Phi_0} b_3
 +\cos (\theta/2) e^{-i k_2 \Phi_0} b_4 =0
 \end{eqnarray}
 \begin{eqnarray}
&& k a_1 -k a_2 +A_1(k_1)e^{i 2\pi k_1} b_1
  +A_1(k_2)e^{i 2\pi k_2} b_2 \nonumber \\
  &&-A_2(k_1)e^{ -i 2\pi k_1} b_3
  -A_2(k_2)e^{ -i 2\pi k_2} b_4  =0
  \end{eqnarray}
  \begin{eqnarray}
&& -k a_3 + k a_4 - A_2(k_1)e^{i 2\pi k_1} b_1
  -A_2(k_2)e^{i 2\pi k_2} b_2 \nonumber \\
 && -A_1(k_1)e^{ -i 2\pi k_1} b_3
  -A_1(k_2)e^{ -i 2\pi k_2} b_4  =0
  \end{eqnarray}
  \begin{eqnarray}
&& ke^{ik\Phi_0} a_1 -k e^{-ik\Phi_0} a_2
 +A_1(k_1)e^{i k_1 \Phi_0} b_1
  +A_1(k_2)e^{i k_2 \Phi_0} b_2 \nonumber \\
 && -A_2(k_1)e^{ -i (k_1+1) \Phi_0} b_3
  -A_2(k_2)e^{ -i (k_2+1) \Phi_0} b_4  =0
  \end{eqnarray}
  \begin{eqnarray}
&& -k e^{-ik\Phi_0} a_3 + k e^{ik\Phi_0} a_4
  - A_2(k_1)e^{i (k_1+1) \Phi_0} b_1
  -A_2(k_2)e^{i (k_2+1)\Phi_0} b_2 \nonumber \\
  &&-A_1(k_1)e^{ -i k_1 \Phi_0} b_3
  -A_1(k_2)e^{ -i k_2 \Phi_0} b_4  =0
\end{eqnarray}
where $A_1(x) \equiv x \cos(\theta/2) -\frac{1}{2} \tan(\theta)
\sin(\theta/2)$ and $A_2(x) \equiv (x+1)\sin(\theta/2) -\frac{1}{2}
\tan(\theta) \cos(\theta/2)$. The eigenvalue $E_n$ can be solved
numerically by setting the determinant of the coefficient of the
variables $a_i$ and $b_i$ in the above eight series of linear
equations to zero.

Now we present the numerical results. Fig.3a shows the eigen values
$E_n$ versus the Rashba SOI's strength $\alpha_R$. For the normal
ring ($\alpha_R =0$), the eigenvalues are $n^2 E_a$ with fourfold
degeneracy, and the corresponding eigenstates are $(1,0)^{\dagger}
e^{\pm i n \varphi}$ and $(0,1)^{\dagger} e^{\pm i n
\varphi}$.\cite{note5} As the SOI is turned on the degenerate energy
levels split while maintaining twofold Kramers degeneracy. The
higher the energy level, the larger this energy split. Typically,
the splits are on the order of $E_a$ at $\alpha_R=10^{-11}eVm$, with
$E_a \approx 0.42meV$ for the ring's radius $a=50nm$ and the
effective mass $m=0.036m_e$. The eigenvalues $E_n$ versus the normal
region's angle $\Phi_0$ are also shown (see Fig.3b). For $\Phi_0
=2\pi$, the whole ring is normal and $E_n$ are fourfold
degenerate.\cite{note5} When $\Phi_0$ is away from $2\pi$, the
degenerated levels are split into two, and the splits are larger
with the smaller $\Phi_0$. When $\Phi_0 =0$, the whole ring has the
Rashba SOI, and the split reaches the maximum.

Since $E_n$ is twofold degenerate, we obtain two eigenstates for
each $E_n$, which are labeled $\Psi_n(\varphi)$ and
$\hat{T}\Psi_n(\varphi)$.\cite{note6} With the wave functions, the
spin current contributed from the level $n$ can be calculated
straightforwardly using the conventional definition $I^n_{S\varphi
i}(\varphi) = Re \Psi_n^{\dagger} \hat{v}_{\varphi} \hat{\sigma}_i
\Psi_n$ ($i=x,y,z$). Notice that the spin current $I_{Sji}$ is a
tensor, where $i,j=r,\varphi,z$ in the cylinder coordinates and
$i,j=x,y,z$ in the orthogonal coordinates. The first index $j$
describes the direction of the motion of the electron, and the
second index $i$ represents the direct of the spin. Because the
device in the present paper is a ring, the motion of the electron
has no components along radial ($r$) and $z$ axis. Hence the spin
currents $I^n_{Sri}$ and $I^n_{Szi}$ ( $I^n_{S(r/z)i}(\varphi) = Re
\Psi_n^{\dagger} \hat{v}_{r/z} \hat{\sigma}_i \Psi_n$) are zero.
Only spin current $I^n_{S\varphi i}$ with the electron moving along
the $\varphi$ direction is non-zero. To simplify the notation,
hereafter we use the symbol $I^n_{Si}$ to replace $I^n_{S\varphi
i}$. Since there is a controversy about the definition of spin
current in the SOI's region, we will calculate the spin current only
in the normal region in this section. The spin current in the SOI's
region will be studied in the section VII after the definition of
the spin current is investigated in the section VI. In the normal
region, the spin current is conserved, so $I^n_{Si}(\varphi)$ is
independent of the angle coordinate $\varphi$.

Fig.4 shows the spin current $I^n_{Si}$ versus the Rashba SOI's
strength $\alpha_R$ for $\Phi_0 =\pi$.\cite{note7} Since $E_n$ is
twofold degenerate, the wave-functions can be arbitrary combination
of $\Psi_n(\varphi)$ and $\hat{T}\Psi_n(\varphi)$. But the spin
current $I^n_{Si}$ remains the same. Our results in Fig.4 show that
$I^n_{Sx}$ is exactly zero for all level $n$ while $\Phi_0 =\pi$,
and $I^n_{Sy}$ and $I^n_{Sz}$ exhibit the oscillatory pattern with
$\alpha_R$. A $\pi/2$-phase shift between $I^n_{Sy}$ and $I^n_{Sz}$
is observed with $\sqrt{(I^n_{Sy})^2 +(I^n_{Sz})^2}$ approximately
constant. For two adjacent levels $2n-1$ and $2n$, their spin
current have opposite signs, and $I^{2n-1}_{Si} + I^{2n}_{Si}=0$ if
$\alpha_R=0$. We note that the spin current $I^n_{Si}$ is quite
large. For example, the value $E_a $ is equivalent to the spin
current of a moving electron in the ring with its speed $4\times
10^5 m/s$.

The spin current $I^n_{Si}$ versus the angle $\Phi_0$ that describes
the normal region at a fixed $\alpha_R =3\times 10^{-11} eVm$ is
shown in Fig.5. When $\Phi_0 =2\pi$, the whole ring is normal,
$I^n_{Sy}$ and $I^n_{Sz}$ are exactly zero. However, $I^n_{Sx}$ is
non-zero (see Fig.5a). Note that $I^{2n-1}_{Sx} + I^{2n}_{Sx}=0$
($n=1,2,3,...$) and $I^0_{Sx}=0$ at $\Phi_0 =2\pi$, so the total
spin current $I_{Sx}$ is still identically zero because that the
(2n-1)-th and the 2n-th states are degenerate and have the same
occupied probability at $\Phi_0 =2\pi$. When $\Phi_0 \not= 2\pi$
where part of the ring has the SOI, three components $I^n_{Sx/y/z}$
of the spin current can be non-zero. For the larger n, the absolute
value of the spin current $|I^n_{Si}|$ is larger. For two adjacent
levels $2n-1$ and $2n$, their spin current have opposite signs,
which is similar to the result of Fig.4.

Now we calculate the equilibrium total spin current $I_{Si}$
contributed from all occupied energy levels: $I_{Si} = 2\sum_n
I^{n}_{Si} f(E_n)$, where $f(E)=1/\{\exp[(E-E_F)/k_B T]+1\}$ is the
Fermi distribution with the Fermi energy $E_F$ and the temperature
$T$ and the factor $2$ is due to the Kramers degeneracy. The
persistent charge current and the equilibrium spin accumulation are
found to be zero because the system has the time-reversal symmetry.
Fig.6a,b show the total spin currents $I_{Si}$ versus the Rashba
SOI's strength $\alpha_R$ for different Fermi energy $E_F$. One of
the main results is that the spin current indeed is non-zero when
$\alpha_R \not=0$. The persistent spin currents $I_{Si}$ in Fig.6
have the following features. At $\alpha_R =0$ the whole ring is
normal, so $I_{Si}$ is exactly zero. With increasing $\alpha_R$,
$I_{Si}$ increases initially and then oscillates for the large
$\alpha_R$. In Fig.6a,b, the parameter $\Phi_0$ is $\pi$, i.e. half
of the ring is normal and the other half of the ring has the SOI,
then $I_{Sx}$ is zero, and only $I_{Sy}$ and $I_{Sz}$ are non-zero.
If $\Phi_0 \not= \pi$, the components $I_{Sx/y/z}$ can be non-zero.
At certain $\alpha_R$, there is a jump in the curve of $I_{Si}$
versus $\alpha_R$. This is because for this $\alpha_R$ the Fermi
energy $E_F$ is in line with a level $E_n$, leading to a change of
its occupation. At zero temperature, the jump is abrupt as shown in
Fig.6a,b. But at finite temperature, this jump will be washed out.
In fact, these results are similar to the persistent (charge)
current in the mesoscopic ring.\cite{aref28}

The spin current $I_{Si}$ versus the angle $\Phi_0$ of normal region
at a fixed $\alpha_R =3\times 10^{-11} eVm$ is shown in Fig.6c,d.
When $\Phi_0 =2\pi$, the whole ring is normal, we have
$I_{Sx/y/z}=0$. When $\Phi_0$ is away from $2\pi$, the spin current
$I_{Sx/y/z}$ emerges. For some Fermi energy a jump appears in the
curve $I_{Si}$-$\Phi_0$ (as shown in Fig.6c), which behaviors is
similar as the jump in the curve $I_{Si}$-$\alpha_R$. For other
Fermi energies, however, the jump in the curve $I_{Si}$ versus
$\Phi_0$ (see Fig.6d) disappears when the Fermi energy $E_F$ is not
in line with the level $E_n$ at all values of $\Phi_0$. In
particular, in the limit when $\Phi_0$ goes to zero, i.e. when there
is no normal region in the ring, the spin current $I_{Sx}$ and
$I_{Sz}$ still exist. This indicates that the normal region is not
necessary for generating $I_{Si}$.

In the above numerical calculation, the temperature $T$ is set to
zero. Now we consider the effect of the finite temperature $T$.
Fig.7 shows the persistent spin currents $I_{Sy/z}$ versus the
strength of SOI $\alpha_R$ at $\Phi_0 =\pi$ with the different
temperatures $k_BT$. When the temperature $k_B T=0$, the spin
currents $|I_{Si}|$ are the largest and $I_{Si}$ shows a jump if the
Fermi energy $E_F$ is in line with a level $E_n$. With the increase
of the temperature from zero, this jump is smoothed, and the spin
currents $|I_{Si}|$ at these $\alpha_R$ near the point of the jump
(e.g. $7.4\times 10^{-11}eVm$ in Fig.7) decreases sharply even at
the very low temperature. But the spin currents $|I_{Si}|$ at these
$\alpha_R$ that are far away from this point of jump (e.g. $\alpha_R
< 5\times 10^{-11}eVm$ in Fig.7) are not effected very much by the
temperature $k_B T$ even for $k_B T$ reaching $0.2 E_a$. Upon
further the temperature $k_B T$ is raised and on the same order of
the energy-level interval, the spin currents $I_{Si}$ is reduced for
all values $\alpha_R$. This is because the probability of
occupation, i.e. $f(E_n)$, of the energy level vary smoothly versus
the level index $n$ and the spin current due to the adjacent levels
are opposite in sign. However, even when $k_B T$ reaching $k_BT =0.5
E_a$, the persistent spin current $I_{Si}$ is still quite large. For
the ring's radius $a=50nm$ and the effective mass $m=0.036m_e$,
$E_a$ is about $0.42meV$. Then the temperature $T$ is approximately
$2.1K$ at $k_BT =0.5 E_a$. This temperature $T$ is easily reached at
the present technology.\cite{aref7,aref45}

\section{Discussion the effect of sharp interface between the SOI region and the
normal region} \label{sec4}

In the above section, the coefficient Rashba SOI $\alpha_R(\varphi)$
vary sharply in the interface of the normal and SOI's part. Now we
examine the effect of sharp interface. Let us consider a hybrid ring
device with the SOI coefficient $\alpha_R(\varphi)$ varying
continuously along the ring. The Hamiltonian is same as Eq.(1), with
$\alpha_R(\varphi)=0$ for $0<\varphi<\pi$ and $\alpha_R(\varphi)=
\alpha_R \sin^2(\varphi)$ for $\pi<\varphi<2\pi$. In this case, both
$\alpha_R(\varphi)$ and $d\alpha_R(\varphi)/d\varphi$ are continuous
at the interfaces $\varphi=0$ and $\pi$. In this section, we show
that persistent spin current still exists for a hybrid ring device
with SOI varying continuously along the ring.

For this continuous varying Rashba SOI coefficient
$\alpha_R(\varphi)$, the Hamiltonian can not be solved analytically.
Here we numerically solve this Hamiltonian by using the discrete
tight-binding model. Notice that $ -\frac{i\sigma_{r}}{2a}\left[
  \alpha_R(\varphi) \frac{\partial}{\partial \varphi}
 + \frac{\partial}{\partial \varphi} \alpha_R(\varphi)
 \right]
  -i \alpha_R(\varphi)\sigma_{\varphi}/(2a)
 = -\frac{i}{2a}\left[
  \alpha_R(\varphi) \sigma_{r} \frac{\partial}{\partial \varphi}
 + \frac{\partial}{\partial \varphi} \sigma_{r} \alpha_R(\varphi)
 \right] $,
and the discretized Hamiltonian discrete becomes,\cite{book1}
\begin{eqnarray}
 &&H = \sum\limits_{j=1}^{N} \left( 2t a_j^{\dagger} a_j - t
 a_{j+1}^{\dagger} a_j -t a_j^{\dagger} a_{j+1} \right) \nonumber
 \\
  && +  \sum\limits_{j=1}^N \left(a_{j+1}^{\dagger} \frac{i}{2a}
  \frac{\sigma_{r,j}\alpha_{R,j} +
  \sigma_{r,j+1}\alpha_{R,j+1}}{2{\Delta\varphi}} a_j +H.c. \right)
\end{eqnarray}
where $a_j$ and $a_j^{\dagger}$ are annihilation and creation
operators at the point $j$, $N$ is the number of the points in the
ring, ${\Delta\varphi}=2\pi/N$ is the angle between two neighboring
points, $t= E_a/({\Delta\varphi})^2$, $\sigma_{r,j}
=\sigma_{r}(j{\Delta\varphi})= \sigma_x \cos(j{\Delta\varphi})
+\sigma_y \sin(j{\Delta\varphi})$, and
$\alpha_{R,j}=\alpha_R(j{\Delta\varphi})$. In the above Hamiltonian,
the point index $N+1$ is same with the point index $1$. Then by
calculating the eigen-values and eigen-vectors of the Hamiltonian
matrix with the dimension $N$, the eigen-values $E_n$ and the eigen
wave-functions $\Psi_{n,j}$ ($\Psi_{n,j}=\Psi_{n}(j\Delta\varphi)$)
of the ring device can be easily solved.\cite{note6} After solving
the eigen wave-functions $\Psi_{n,j}$, the spin current
$I_{Sk}^n(\varphi) = Re \Psi_n^{\dagger}
\hat{v}_{\varphi}\hat{\sigma}_k \Psi_n = \frac{\hbar^2}{2ma}Re(-i)
\Psi_n^{\dagger} \hat{\sigma}_k \frac{\partial}{\partial \varphi}
\Psi_n $ ($k=x,y,z$) in the normal region can be obtained
straightforwardly from
\begin{equation}
  I_{Sk,j}^n =  \frac{\hbar^2}{2ma}Im \Psi_{n,j}^{\dagger} \hat{\sigma}_k
 \frac{ \Psi_{n,j+1}-\Psi_{n,j-1}}{2{\Delta\varphi}},
\end{equation}

In order to show that the above method is correct and reliable, we
first solve the model of the section III again, with the sharp
varying SOI's coefficient $\alpha_R(\varphi)$: $\alpha_R(\varphi)=0$
while $0<\varphi<\Phi_0$ and $\alpha_R(\varphi)=\alpha_R$ while
$\Phi_0 < \varphi < 2\pi$. The results of the eigen-energies are
shown in Fig.8. When the number of points $N=20$, we can see that
the eigen-energies from the above discrete method are quite
different from the exact values obtained from the method in the
section III (see Fig.8a). But with increasing $N$ (e.g. $N=50$),
this difference become very small (see Fig.8b)). When $N=150$, the
eigen-energies from the discrete method are in excellent agreement
with the exact results (see Fig.8c). This means that our results
using the above discrete method converges for large $N$.

Since $E_n$ is twofold degenerate, the arbitrary combination
$c_1\Psi_n(\varphi) +c_2 \hat{T}\Psi_n(\varphi)$ still is the eigen
wave-functions, so the wave-function is un-certain. In the following
we examine the correctness of our spin current $I_{Si}^n$
($i=x,y,z$), which is calculated from the wave-functions. Fig.9
shows the spin currents $I^n_{Sy}$ contributed from the level $n$
for the case of the sharp varying SOI, and it shows that the results
of the spin current of the above discrete method are also in
excellent agreement with the exact result at large $N$. In
particular, it is surprising that the number of points $N$ need not
to be very large. For $N=50$, the difference between numerical and
the exact results is already very small, and for $N=150$, there is
almost no difference.

Now we are ready to examine the effect of sharp interface. Fig.10
shows the results of the hybrid ring with SOI varying continuously
along the ring, with $\alpha_R(\varphi)=0$ for $0<\varphi<\pi$ and
$\alpha_R(\varphi)= \alpha_R \sin^2(\varphi)$ for $\pi<\varphi<2\pi$
otherwise. Fig.10a, b, and c are for the eigen values, the spin
current $I^n_{Sy}$ contributed from the level $n$, and the
equilibrium total spin current $I_{Si}$ in the normal region versus
$\alpha_R$, respectively. The eigen values are four-fold degenerate
at $\alpha_R=0$, and this degeneracies are split into two twofold
Kramers degenerated states. The spin current $I^n_{Sy}$ contributed
from the level $n$ oscillates with $\alpha_R=0$. These results of
the eigen values and $I^n_{Sy}$ are similar to case of the sharp
interface. In particularly, the results show that the (equilibrium)
persistent spin current $I_{Si}$ is still non-zero and has a quite
large value (see Fig.10c, d). This indicates that the persistent
spin current $I_{Si}$ indeed is originated from the SOI, and it is
not the artifact of sharp interface.

\section{relation between the persistent spin current and the transport spin current}
 \label{sec5}

Through the physical arguments and physical pictures from four
different aspects in the section II, and the analytic results of an
example of a SOI-normal ring (in which the definition of the spin
current is without controversy in the normal region) in the section
III, as well as the discussion concerning the interface in the
section IV, so far we have plenty evidence to demonstrate that the
persistent spin current indeed exists in a mesoscopic semiconducting
ring device with an intrinsic SOI. This persistent spin current can
be induced solely by a SOI, and it exists in an equilibrium
mesoscopic device without a magnetic field, a magnetic flux, and in
the absence of any magnetic materials. Besides the above mentioned
ring geometry, we find that the device can also be of other
shapes,\cite{aref39,aref47} e.g., a disc device, a quantum wire, a
two-dimensional system, {\it etc}. Thus, it is a generic feature
that a pure persistent spin current appears in a system with SOI. In
this section, we will discuss the relation between the persistent
spin current and the normal transport spin current.

First let us review and discuss the relation of the persistent
(charge) current and the transport (charge) current in a mesoscopic
ring device. To consider a mesoscopic ring threaded by a magnetic
flux coupled to two leads (the lead-L and the lead-R) which act as
the electron reservoirs (see Fig.11a). There exists dissipation in
the two leads and they are always in equilibrium for an isolated
lead-L(R). The size of the ring is assumed within the coherent
length and no dissipation in the ring. At zero bias (i.e. the
equilibrium case), the transport charge current is zero everywhere,
including the two leads and the ring, but the persistent charge
current exists only in the ring, but not the leads because of
presence of dissipation in the leads. If a non-zero bias is added
between two leads, a transport charge current flows from one lead
through the ring to the other lead, so there exists both transport
and persistent charge currents in the ring but only the transport
current in the leads because of dissipation. In other words, at a
finite bias, the charge currents $I_{e1(e2)}$ in the two arms of the
ring are the sum of the transport currents $I_{e1(e2),t}$ and the
persistent current $I_{e,p}$: $I_{e1}= I_{e1,t}+ I_{e,p}$ and
$I_{e2}= I_{e2,t}- I_{e,p}$. Can one distinguish the transport
charge currents $I_{e1(e2),t}$ and the persistent charge current
$I_{e,p}$? In fact, they ($I_{e1(e2),t}$ and $I_{e,p}$) can not be
distinguished either in theory or in experiment. The transport
charge currents $I_{e1(e2),t}$ and the persistent charge current
$I_{e,p}$ in the ring have identical behaviors, both of them are
dissipationless\cite{note8}, capable of inducing a magnetic field,
{\it etc}. So in principle only the total charge currents
$I_{e1(e2)}$ in the arms are observable physical quantities. If some
dissipative impurities are introduced in the ring, then the
transport charge currents $I_{e1(e2),t}$ show dissipation while the
persistent charge current $I_{e,p}$ does not. Under this
circumstance, can one distinguish the transport currents
$I_{e1(e2),t}$ and the persistent currents $I_{e,p}$? It turns out
that one can still not distinguish these types of currents. When
dissipative impurities are introduced, the original dissipationless
ring is changed into a new different ring, in which the persistent
charge current $I_{e,p}$ is quenched while the transport charge
currents $I_{e1(e2),t}$ are normally reduced due to dissipation.
Thus, it is impossible to obtain $I_{e1(e2),t}$ and $I_{e,p}$ of the
original dissipationless ring.

The relation between the persistent spin current and the transport
spin current is identical to the relation between the two charge
currents discussed above. Consider that part of a mesoscopic ring
contains a SOI but without a magnetic flux, and with two leads
coupled to this ring (see Fig.11b). In the equilibrium case, a
persistent spin current emerges in the ring, neither the transport
spin current nor the persistent spin current is present in the
leads. Under a spin-motive force,\cite{aref18,aref49,aref50} a
transport spin current flows from one lead through the ring to the
other lead. The persistent spin current in the leads is always
absent independent of with or without a spin-motive force due to
dissipation. On the other hand, with a spin motive force, both
persistent spin current and the transport spin current exist in the
ring. Similarly to the charge currents, both persistent spin current
$I_{s,p}$ and transport spin currents $I_{s1(s2),t}$ in the two arms
of the ring are indistinguishable since they behave identically in
all physically measurable properties. Both of them are
dissipationless\cite{note8}, describing the real spin motion,
capable of inducing an electric field, and so on.

\section{the definition of the spin current}
\label{sec6}

From the sections II, III, and IV, we have made sure that the
persistent spin current exists in the equilibrium mesoscopic
semiconducting device in the presence of SOI. In this section, we
examine the definition of the spin current. The first work to
question the conventional definition of the spin current is by
Rashba.\cite{aref23} After that, many subsequent papers have
discussed the definition of the spin current as mentioned in the
introduction.\cite{aref24,aref25,aref26,aref27} In
summary,\cite{aref26} in the presence of SOI one faces three
problems when using the conventional definition ${\bf I}_S = Re
\Psi^{\dagger} \hat{v} \hat{s} \Psi$:\cite{note2} (i) there exists
a non-zero spin current even in the equilibrium system, (ii) the
spin current usually is non-conservative, and (iii) the Onsager
relation is violated. Therefore suggestions have been made in
previous papers that one needs to modify the conventional
definition of the spin current. In the following, we examine these
three problems in detail and we argue that there is no need to
modify this conventional definition\cite{note2}  ${\bf I}_S = Re
\Psi^{\dagger} \hat{v} \hat{s} \Psi$.

\begin{center}
{\bf (i) the non-zero spin current in the equilibrium system:}
\end{center}

From the investigation and the discussion in the sections II, III,
and IV, we have clearly shown that this non-zero spin current is the
persistent spin current. It describes the real motion of spins, has
the physical meaning, and can be observed in the experiment in
principle (see the section VIII). So this non-zero spin current
should be kept in the calculation of the total spin current.

\begin{center}
{\bf (ii) non-conservation of the spin current:}
\end{center}

In this sub-section, we argue that in general the spin current is
not conserved in the presence of the SOI or/and the magnetic field.
However, in certain cases, the spin current can be conserved by
including the strong spin-spin interaction. Let us analyze this
problem from the both aspects of the theory and experiment. In the
aspect of the theory, we give two examples of the non-conserved spin
current (for intuition the readers can also consider the spin as a
classic vector).

First, we want to show that in general if the system has spin flip
mechanism, the spin current will not be conserved. For instance, in
the presence of a rotating magnetic field or circular polarized
light\cite{aref51} the spin current exists in a device connected to
only one terminal. This one terminal device and its spin
translational motion and precession are as shown in Fig.12a. Here at
$x<0$, the spin is without precession, and the spin pointing to the
$y$-direction moves along the $+x$-direction while the spin pointing
to the $-y$-direction moves along the opposite direction. So it has
a non-zero spin current $I_{s,xy}$ at $x<0$. Near the point $x=0$
where the quantum dot located, the spin precesses and is flipped due
to the presence of a rotating magnetic field or circularly polarized
light.\cite{aref51} Then the spin accumulation does not vary with
time and the system maintains the steady state. It is obvious that
the spin current is not conserved because of only one terminal in
this device.

In the second theoretical example, to consider the spin
translational motion and precession on a ring as shown in Fig.12b.
At point A with its angular coordinate $\varphi=0$, the spin
pointing to $+x$-direction moves down while the spin pointing to
$-x$-direction moves up, then the non-zero element of the spin
current is $I_{S,yx}$. At another point B with its angular
coordinate $\varphi=\pi/2$, the spin pointing to $+y$-direction
moves along the $+x$-axis and the spin pointing to $-y$-direction
moves along the $-x$-axis, then the non-zero element of the spin
current is $I_{S,xy}$. So it is obvious that the spin current is not
conserved in this ring device, but the spin accumulation still keep
invariant. In fact, this example is similar to the persistent spin
current in the mesoscopic ring while the whole ring has the SOI (see
the section VII.C and the appendix), in which the motion of the
$x$-$y$- plane elements of the spin is as in the Fig.12b and the
spin accumulation is zero everywhere.

In addition, there are experimental indications that the spin
current is not conserved. For example, let us consider the
experimental result by Valenzuela and Tinkham.\cite{aref21} In this
experiment, they have clearly shown that a pure spin current injects
from the FM1 electrode into the Al strip, reduces with its flowing
forward because of the spin flip, and finally disappears while the
distance much longer than the spin diffusion length (see Fig.1a, b,
c, and Fig.4 in Ref.\cite{aref21}). This experimental result gives a
strong proof of that the spin current should be non-conserved in the
presence of the SOI, the magnetic field, the magnetic impurity, or
others.

\begin{center}
{\bf (iii) violation of the Onsager relation:}
\end{center}

The Onsager reciprocal relation is an important theorem of the
near-equilibrium transport theory. Up to now, the Onsager reciprocal
relation is always satisfied for transport of any physical quantity
(e.g. the charge transport, thermal transport, etc) by suitably
defining a corresponding current. However, in the following we point
out that the Onsager relation for the spin transport is in general
violated. In particular, it is impossible to restore the Onsager
relation regardless of how to modify the definition of the spin
current. This is very different from all previous cases. In this
sub-section, we first recall the tenable condition of the Onsager
reciprocal relation, and then examine the case of the spin current.
We find that the spin transport does not always meet this condition
no matter how to define the spin current, although this condition is
met for all previous transport phenomena people have studied. So the
Onsager relation is in general violated for the spin transport.

First, let us recall the tenable condition of the Onsager relation.
Considering the currents $\vec{I}=\{I_i\}$ and its corresponding
forces $\vec{F}=\{F_i\}$, and they have the following relationships:
\begin{equation}
  I_i =\sum\limits_{j} G_{ij} F_j,
  \label{eq14}
\end{equation}
where $G_{ij}$ is the conductivity. {\sl If} the local entropy
production $dS/dt$ per unit time can be expressed as: $dS/dt= \sum_i
I_i F_i$, there exists an Onsager relation $G_{ij} = G_{ji}$
(assuming that the system has the time-reversal symmetry).

Next, we give an intuitive example to show the tenable condition of
the Onsager relation. Let us consider the charge conductivity in the
two-dimensional system. To take the vectors $\vec{e}_x$ and
$\vec{e}_y$ (as shown in Fig.13a) as the base vectors of the charge
current and its force [i.e. the gradient of the potential,
$\vec{\nabla} V({\bf r}) $], the equation (\ref{eq14}) becomes:
\begin{eqnarray}
 \left( \begin{array}{l} I_x \\I_y \end{array}\right) =
 \left( \begin{array}{ll} G_{xx} & G_{xy} \\ G_{yx} & G_{yy} \end{array}\right)
 \left( \begin{array}{l} \nabla_x V \\ \nabla_y V
 \end{array}\right).
  \label{eq15}
\end{eqnarray}
In this case, the Onsager relation is tenable, and
$G_{xy}=G_{yx}\equiv G_o$. However, if taking the non-orthonormal
vectors $\vec{e}_x$ and $\vec{e}_{r}$ (as shown in Fig.13b) as the
base vectors of the current and the force, the Onsager relation will
be violated as shown in the following. For the base vectors
$\vec{e}_x$ and $\vec{e}_{r}$, any current vector $\vec{I}$ and
force vector $\vec{\nabla} V$ can still be expressed as: $\vec{I} =
I_x \vec{e}_x + I_{r} \vec{e}_{r}$ and $\vec{\nabla} V = \nabla_x V
\vec{e}_x + \nabla_{r} V \vec{e}_{r}$. Then it is easily to obtain
the relation between $(\nabla_x V, \nabla_y V)$ and $(\nabla_x V,
\nabla_{r} V)$:
\begin{eqnarray}
 \left( \begin{array}{l} \nabla_x V \\ \nabla_{y}V \end{array}\right) =
 {\bf U} \left( \begin{array}{l} \nabla_x V \\ \nabla_r V \end{array}\right)
 ,
 \hspace{3mm} and
 \left( \begin{array}{l} \nabla_x V\\ \nabla_r V\end{array}\right) =
 {\bf U}^{-1} \left( \begin{array}{l} \nabla_x V \\ \nabla_{y}V
 \end{array}\right),
 \nonumber
\end{eqnarray}
where
\begin{eqnarray}
 {\bf U}=
\left( \begin{array}{ll} 1 & \cos\varphi \\
0  & \sin\varphi
 \end{array}\right)
\end{eqnarray}
So under the base vectors $\vec{e}_x$ and $\vec{e}_{r}$, the
Eq.(\ref{eq15}) changes to:
\begin{eqnarray}
 \left( \begin{array}{l} I_x \\I_{r} \end{array}\right)  & = &
  {\bf U}^{-1}\left( \begin{array}{ll} G_{xx} & G_o \\ G_o & G_{yy} \end{array}\right)
  {\bf U}
 \left( \begin{array}{l} \nabla_x V \\ \nabla_{r} V
 \end{array}\right) \nonumber \\
 & = &   \left( \begin{array}{ll} G_{xx}
-G_o\cos\varphi/\sin\varphi &
  (G_{xx}-G_{yy})\cos\varphi-G_o\cos(2\varphi)/\sin\varphi
 \\
 G_o/\sin\varphi  &
 G_{yy}+G_o\cos\varphi/\sin\varphi
\end{array}\right)
 \left( \begin{array}{l} \nabla_x V \\ \nabla_{r} V
 \end{array}\right).
\end{eqnarray}
It is obvious that the two off-diagonal elements of the conductivity
in the above equation are not equal, so the Onsager relation is
violated in the non-orthonormal base vectors $\vec{e}_x$ and
$\vec{e}_{r}$. In fact, for the charge current, the Onsager relation
is only tenable under the orthonormal and linear independent base
vectors.

Now, let us discuss the spin current. The spin current has $3\times
3=9$ elements and the charge current has $3$ elements. So here the
current $\vec{I}$ (including spin and charge) and the corresponding
force $\vec{F}$ totally has $12$ elements and the conductivity has
$12\times 12=144$ elements. For simplicity and clarity, we consider
a one-dimensional system and the electron can only move along the
$x$ axis. In this case, the current $\vec{I}$ and the force
$\vec{F}$ have only 4 non-zero elements, and they are:
\begin{eqnarray}
 \vec{I} & = & (I_{sxx},I_{sxy}, I_{sxz}, I_{cx}),\\
 \vec{F} &=& (\nabla_x V_{sx}, \nabla_x V_{sy},\nabla_x V_{sz},\nabla_x
 V_{c}),
\end{eqnarray}
where $V_{si}$ ($i=x,y,z$) is the spin chemical potential and
$V_{c}$ is the (charge) chemical
potential.\cite{aref18,aref49,aref50} The spin chemical potential
$V_{si}$ means that the electron of the spin pointing to $+i$ and
$-i$-direction occupy up to $V_c +V_{si}/2$ and $V_c - V_{si}/2$
(shown in Fig.13c, d, e), respectively. Then the Eq.(\ref{eq14})
becomes:
\begin{eqnarray}
\left( \begin{array}{l} I_{sxx} \\ I_{sxy} \\ I_{sxz}
 \\ I_{cx} \\ \end{array}\right)
 =
 \left( \begin{array}{llll} G_{xx} & G_{xy}& G_{xz}& G_{xc} \\
   G_{yx} & G_{yy}& G_{yz}& G_{yc} \\
   G_{zx} & G_{zy}& G_{zz}& G_{zc} \\
   G_{cx} & G_{cy}& G_{cz}& G_{cc}
  \end{array}\right)
 \left( \begin{array}{l} \nabla_{x} V_{sx} \\ \nabla_{x} V_{sy} \\
  \nabla_{x} V_{sz} \\\nabla_{x} V_{c}
 \end{array}\right)
 \label{eq20}
\end{eqnarray}
and the conductivity has $4\times 4 =16$ elements.

Now we show that the spin transport does not meet the tenable
condition of the Onsager relation. First, the three base vectors
$\sigma_i$ ($i=x,y,z$) are not orthonormal, and it is impossible to
find a series of the orthonormal base vectors regardless of what
combination of the three $\sigma_i$.

In addition, once the spin chemical potential of one component is
fixed, the other two spin potentials $V_{si}$ can not exist. For
example, if we give the value for the spin potential $V_{sz}$, this
means that the electron of the spin along $+z$ direction occupies up
to $V_{sz}/2$ and the electron of the spin at $-z$ direction
occupies up to $-V_{sz}/2$ if $V_c=0$ (see
Fig.13e),\cite{aref18,aref49} and the electron occupational state
has completely been determined. Then one can not further specify the
spin states and the corresponding occupation number along $+x$ and
$-x$ (or $+y$ and $-y$). Therefore, while giving the value for one
component of the spin potential, e.g. $V_{sz}$, the others $V_{sx}$
and $V_{sy}$ do not exist. This conclusion hold even for spin free
and conserved system with $[\sigma_i, H]=0$ for all $i$ ($i=x,y,z$).
Due to the fact that three spin potentials $V_{si}$ can not be
simultaneously evaluated, the Eq.(\ref{eq20}) [i.e. the
Eq.(\ref{eq14})] does not exist for the spin transport, regardless
of how to define the spin current. Therefore there is no Onsager
relation in spin transport, and it is impossible to restore the
Onsager relation by modifying the definition of the spin current. In
fact, the Onsager relation is not satisfied for any existing
definitions of the spin current.

Furthermore, we can also use the four normal terminal device (as the
device in Ref.\cite{aref18}) to examine the Onsager relation.
Because the four terminal leads are normal metal without the SOI,
there is no spin-flip. For this system, there is no controversy for
the definition of the spin current in the terminal leads. Therefore
we can avoid the definition of the spin current and examine the
Onsager relation. The results also show that the Onsager relation
does not exist for the spin transport. This gives an additional
proof.

We wish to mention that because $V_{sz}$ and $V_c$ (or $V_{sx}$ and
$V_c$, or $V_{sy}$ and $V_c$) can be determined simultaneously, the
Onsager relation for $G_{zc} = G_{cz} $ (or $G_{xc} = G_{cx} $, or
$G_{yc} = G_{cy}$) might exist,\cite{aref18,aref26} e.g., in the
suitable boundary condition in the four terminal
device.\cite{aref18} However, it is impossible that six relations
($G_{zc} = G_{cz} $, $G_{xc} = G_{cx} $, $G_{yc} = G_{cy}$, $G_{xy}
= G_{yx}$, $G_{xz} = G_{zx}$, and $G_{yz} = G_{zy}$) are satisfied
simultaneously.\cite{aref18} For three-dimensional systems, the
conductivity has $12\times 12 =144$ elements $G_{ij}$, similar
conclusion applies, i.e., it is impossible that $G_{ij}=G_{ji}$ are
satisfied for all off-diagonal matrix element simultaneously.

\begin{center}
{\bf (iv) discussions }
\end{center}

From the discussion in the last three sub-sections, we have clearly
shown that (i) the non-zero spin current in the equilibrium SOI's
device is the persistent spin current, (ii) in general the spin
current is not conserved, and (iii) the Onsager relation is violated
for the spin transport. In particular, it can not be restored
regardless of how to modify the definition of the spin current.
Therefore the three "flaws" of the conventional definition of the
spin current, which has been mentioned and commented in some
previous papers, are intrinsic properties of spin transport. In
addition, the conventional definition has very clear physical
intuition and has described the spin motion very well. Using this
conventional definition, one can account for many effects that
relate the spin transport,\cite{aref24} e.g., the heat produced by
the spin current, the spin currents induced electric field, and the
force and the torque acting on the spin current in the presence of
electric field. Therefore we make the conclusion that there is no
need to modify the conventional definition ${\bf I}_s = Re
\Psi^{\dagger} \hat{v} \hat{s} \Psi$.\cite{note2}

Before the end of this section, we wish to mention that if the
spin-spin interaction is included into the Hamiltonian, the spin
current calculated using the conventional definition will be
conserved automatically.\cite{aref27} We will discuss this in detail
in section IX. We also note that our discussion above does not
contradict with the angular spin current in our previous paper by
Sun and Xie.\cite{aref24} To see this, let us first recall the
(linear) velocity $\vec{v}$ and the angular velocity $\vec{\omega}$.
(a) The (linear) velocity $\vec{v} = d{\bf r}/dt$ (or the velocity
operator $\hat{v} = d\hat{\bf r}/dt$) can describe the translational
motion of the vector (or the rigid body) and there is no need to
modify this velocity definition $\vec{v} = d{\bf r}/dt$. (b) On the
other hand, the vector has the rotational degrees of freedom except
for its translational motion, so we need to introduce the angular
velocity $\vec{\omega}$ to describe its rotational motion. We
emphasize that two statements (a) and (b) do not contradict to each
other and can be rephrased as (a) The (linear) spin current ${\bf
I}_s = Re \Psi^{\dagger} \hat{v} \hat{s} \Psi$ can describe the
translational motion of the spin and there is no need to modify its
definition. (b) On the other hand, the spin has the rotational
degrees of freedom (precession) except for its translational motion,
so we need to introduce the angular spin current ${\bf I}_{\omega}
=Re \Psi^{\dagger} (d \hat{s}/dt) \Psi= Re \Psi^{\dagger}
\hat{\omega} \times \hat{s} \Psi$ to describe its rotational
motion.\cite{note9}

\section{the persistent spin current in the region with SOI}
\label{sec7}

After clarifying the definition of the spin current, we return to
discuss the persistent spin current in the SOI's region of the
normal-SOI ring in this section. In the sub-section A, we
investigate the linear and angular persistent spin current. In the
sub-section B, we show that the persistent spin current still exists
even using the new definition of the spin current as in
Ref.[\onlinecite{aref26}]. The case of entire ring with SOI is
studied in the sub-section C. At the last sub-section D, we present
some conserved quantities that are related to the persistent spin
current.

\subsection{the linear and angular persistent spin currents in the SOI's
region}

From the definition of the (linear) spin current ${\bf I}_{Si} =
Re \Psi^{\dagger} (1/2)(\hat{v}_{\varphi} \hat{s}_i + \hat{s}_i
\hat{v}_{\varphi}) \Psi$, for a ring device we have:
\begin{eqnarray}
 \left\{ \begin{array}{lll}
   I_{Sx}(\varphi) & = & Im \{ \Psi^{\dagger}[\sigma_x a E_a
   \frac{\partial}{\partial \varphi}
   +\frac{i\alpha_R(\varphi)}{2}\cos \varphi] \Psi\} \\
   I_{Sy}(\varphi) & = & Im \{ \Psi^{\dagger}[\sigma_y a E_a
   \frac{\partial}{\partial \varphi}
   +\frac{i\alpha_R(\varphi)}{2}\sin \varphi] \Psi\} \\
   I_{Sz}(\varphi) & = & Im \{ \Psi^{\dagger}[\sigma_z a E_a
   \frac{\partial}{\partial \varphi}]
    \Psi\}
\end{array}\right.
\end{eqnarray}
To make the above equations discrete, $I_{Si}(\varphi)$ changes
into $I_{Si,j}=I_{Si}(j\Delta\varphi)$:
\begin{eqnarray}
 \left\{ \begin{array}{lll}
   I_{Sx,j} & = & a E_a Im \left\{ \Psi_j^{\dagger}\left[\sigma_x
   \frac{\Psi_{j+1}-\Psi_{j-1}}{2\Delta\varphi}
   + \frac{i\alpha_{R,j}}{2E_a a}\cos (j\Delta\varphi)\Psi_j\right]\right\} \\
   I_{Sy,j} & = & a E_a Im \left\{ \Psi_j^{\dagger}\left[\sigma_y
   \frac{\Psi_{j+1}-\Psi_{j-1}}{2\Delta\varphi}
   + \frac{i\alpha_{R,j}}{2E_a a}\sin (j\Delta\varphi)\Psi_j\right]\right\} \\
   I_{Sz,j} & = & a E_a Im \left\{ \Psi_j^{\dagger} \sigma_z
   \frac{\Psi_{j+1}-\Psi_{j-1}}{2\Delta\varphi}
 \right\}
\end{array}\right.
\end{eqnarray}
Let us calculate the angular spin current ${\bf
I}_{\omega}$,\cite{aref24} which describes the rotational motion
(precession) of the spin. For a ring device with the Hamiltonian of
Eq.(1), one has:
\begin{eqnarray}
 \left\{ \begin{array}{lll}
   d\hat{s}_x/dt & =& (-i\sigma_z/2a)
   \{\alpha_R(\varphi)\sin\varphi,\partial/\partial \varphi\}, \\
   d\hat{s}_y/dt & =& (i\sigma_z/2a)
   \{\alpha_R(\varphi)\cos\varphi,\partial/\partial \varphi\}, \\
   d\hat{s}_z/dt & =& (-i/2a)
   \{\alpha_R(\varphi)\sigma_{\varphi},\partial/\partial \varphi\},
\end{array}\right.
\end{eqnarray}
where $\{A,B\}\equiv AB+BA$. Using the cylindrical coordinates,
$d\hat{\vec{s}}/dt =(d\hat{s}_r/dt,
d\hat{s}_{\varphi}/dt,d\hat{s}_z/dt) = \frac{-i}{2a}(0,
-\sigma_z\{\alpha_R(\varphi), \partial/\partial \varphi\},
 \{\alpha_R(\varphi)\sigma_{\varphi},\partial/\partial \varphi\})$.
Then the angular spin current can be calculated from its
definition,\cite{aref24} ${\bf I}_{\omega} =Re \Psi^{\dagger} (d
\hat{\vec{s}}/dt) \Psi$ straightforwardly, so does its discrete
version:
\begin{eqnarray}
 \left\{ \begin{array}{lll}
   I_{\omega r,j} & =& 0 \\
   I_{\omega \varphi,j} & =&
    Re \left\{ \Psi^{\dagger}_j \frac{i\sigma_z}{2a} \left[
    \alpha_{R,j}\frac{\Psi_{j+1} -\Psi_{j-1}}{2\Delta\varphi}
    + \frac{\alpha_{R,j+1}\Psi_{j+1} -\alpha_{R,j-1}\Psi_{j-1}}{2\Delta\varphi}
     \right]\right\}\\
   I_{\omega z,j} & =&
    Re \left\{ \Psi^{\dagger}_j \frac{-i}{2a} \left[
    \sigma_{\varphi,j}\alpha_{R,j}\frac{\Psi_{j+1} -\Psi_{j-1}}{2\Delta\varphi}
    + \frac{\sigma_{\varphi,j+1}\alpha_{R,j+1}\Psi_{j+1}
    -\sigma_{\varphi,j-1}\alpha_{R,j-1}\Psi_{j-1}}{2\Delta\varphi}
     \right]\right\}
\end{array}\right.
\end{eqnarray}

Since the wave function $\Psi_n(\varphi)$ for each eigenstate is
known from the section III or IV, the (linear) spin current
$I^n_{Si}$ and the angular spin current $I^n_{\omega i}$ from the
$n$-th state can be easily obtained. Using the exact method in the
section III or the discrete method in the section IV, we find that
the results for $I^n_{Si}$ and $I^n_{\omega i}$ match perfectly.
Fig.14 shows the linear spin current $I^n_{Sx/y/z}$ (i.e. the
element in the orthogonal coordinates) and the angular spin current
$I^n_{\omega \varphi/z}$ (i.e. the element in the cylindrical
coordinates) versus the angular coordinates $\varphi$. In the normal
region, $I^n_{Sx/y/z}$ is conserved and is independent of the
coordinates $\varphi$, and $I^n_{\omega \varphi/z}$ is zero, since
the spin has only the translational motion without the precession
there. On the other hand, in the SOI region, except for $I^n_{\omega
r} =0$, the (linear) spin current $I^n_{sx/y/z}$ and the angular
spin current $I^n_{\omega \varphi/z}$ are all non-zero and
non-conserved, because of the spin precession in the presence of the
SOI. The linear spin current $I^n_{Sx/y/z}(\varphi)$ versus
$\varphi$ is always continuous, even at the interface between the
normal region and the SOI region where the strength of SOI
$\alpha_R(\varphi)$ changes abruptly. But the angular spin current
$I^n_{\omega \varphi/z}(\varphi)$ versus $\varphi$ shows a jump
whenever $\alpha_R(\varphi)$ versus $\varphi$ has as abrupt change,
as shown in Fig.14d, e, and f. The jump position of $I^n_{\omega
\varphi/z}$ is located at the abrupt point of $\alpha_R(\varphi)$.
The spin currents $I^n_{Si}$ (and $I^n_{\omega}$) versus $\varphi$
for the states with the same-parity, e.g. $n=1,3,5,...$, or
$n=0,2,4,6,...$, are similar in shape (see Fig.14a, c, d, and f),
but the value of $|I^n_{Si}|$ (or $|I^n_{\omega}|$) is much larger
for a larger $n$. Thus, the spin current from the highest occupied
level dominates in the (total) persistent spin current.

Fig.15 shows the total persistent spin current $I_{Sx/y/z}$ versus
the angular coordinates $\varphi$ for different Fermi energies
$E_F$. The persistent spin current is conserved in the normal region
but not so in the SOI region because of the spin precession. Due to
the fact that the spin current $I_{Sx/y/z}^{maxn}$ from the highest
occupied level $maxn$ dominates in $I_{Sx/y/z}$, $I_{Sx/y/z}$
behaves similarly as $I_{Sx/y/z}^{maxn}$ (see Fig.14 and Fig.15).

\subsection{calculating the spin current using the definition in
Ref.[\onlinecite{aref26}]}

We have clarified and demonstrated that the conventional definition
of the spin current makes sense in the section VI. However, in this
sub-section, we present our calculated results of the persistent
spin current by using the new definition of the spin current as
appeared in Ref.{\onlinecite{aref26}}, and discuss its consequences.
In the normal region, the persistent spin current ${\bf I}_S$ is the
same regardless which definition is used. But in the SOI region,
${\bf I}_S$ depends on the definition. In the following let us
discuss ${\bf I}_S$ in the SOI region. By using the new definition
in Ref.{\cite{aref26}}, ${\bf I}_S \equiv Re \Psi^{\dagger} \frac{d
{\bf r} \hat{s}}{dt} \Psi = Re \Psi^{\dagger} \left[\hat{v} \hat{s}
+ {\bf r} d\hat{s}/{dt} \right] \Psi$, the spin current from the
level $n$, then the total persistent spin current, depends on the
choice of the coordinate origin. First, if setting the origin at the
center of the ring, the element $I_{S\varphi i}$ ($i=x,y,z$) with
the spin motion along the $\varphi$ direction is $ Re \Psi^{\dagger}
\hat{v}_{\varphi} \hat{s_i} \Psi$. So $I_{S\varphi i}$ is completely
the same to the result using the conventional definition, and it is
still non-zero in the equilibrium case and non-conversed in the
presence of a SOI. On the other hand, the element $I_{Sr i}$
($i=x,y,z$) with the spin motion along the radial direction is $ a
Re \Psi^{\dagger} (d\hat{s}_{i}/dt) \Psi = a I_{\omega i}$, so it is
also non-zero, but the same element is exactly zero using the
conventional definition. The element $I_{Sz i}$ ($i=x,y,z$) with the
spin motion along the z direction is zero, same as with the
conventional definition. Second, if the coordinate origin is not
located at the center of the ring, all 9 elements of the spin
current are in general different to those by using the conventional
definition. In particular, they are all non-zero and non-conserved
in equilibrium.

\subsection{the case when the entire ring is with a SOI}

Let us consider the case of $\Phi_0 \rightarrow 0$, i.e. the normal
region is gradually getting smaller and at the end the whole ring
contains the SOI. Fig.16 shows the persistent spin current
$I_{Sx/y/z}$ versus the angle coordinate $\varphi$ for $\Phi_0 =
\pi/2$, $\pi/4$, and $0$, respectively. The results clearly show
that the persistent spin current $I_{Sx/y/z}$ does exist, and its
value $|I_{Sx/y/z}|$ is even larger with decreasing of the normal
region, i.e. $\Phi_0$. Eventually when the entire ring contains the
SOI, $|I_{Sx/y/z}|$ reaches its maximum value. This means that the
normal region is not necessary for the existence of the persistent
spin current.

In fact, if the whole ring has a constant SOI with
$\alpha_R(\varphi) =\alpha_R$, the persistent spin current can be
analytically obtained. In this case, the eigenwave function
is:\cite{aref31}
\begin{equation}
 \Psi_n(\varphi) = \left(\begin{array}{l}
       \cos (\theta/2) e^{i n\varphi} \\
       -\sin (\theta/2) e^{i(n+1)\varphi}
       \end{array}
       \right),
       \label{eq25}
\end{equation}
where $n=0,\pm1,\pm2,...$ and the eigenvalue $E_n$ is given by
\begin{equation}
E_n = E_a [n^2 + (n+1/2) (1-1/\cos\theta)]
\end{equation}
Then the linear spin current ${\bf I}^n_S$ and the angular spin
current ${\bf I}^n_{\omega}$ from the state $n$ are as follows:
\begin{eqnarray}
 {\bf I}^n_{S} &= & -E_a F(\theta) \left[\vec{e}_x \sin\theta \cos \varphi +
  \vec{e}_y\sin\theta \sin\varphi -\vec{e}_z \cos\theta  \right] \label{eq27} \\
 {\bf I}^n_{\omega} &= & (E_a/a) F(\theta) \left[ \vec{e}_x\sin\theta \sin\varphi
 -  \vec{e}_y \sin\theta \cos \varphi \right] \nonumber \\
  &=& -(E_a/a) F(\theta) \vec{e}_{\varphi} \sin\theta
  \label{eq28}
\end{eqnarray}
where $ F(\theta)=[n+1/2-1/(2\cos\theta)]/2\pi $. The persistent
linear and angular spin currents ${\bf I}_S$ and ${\bf I}_{\omega}$
are obtained by summing ${\bf I}^n_S$ and ${\bf I}^n_{\omega}$ over
the occupied states. In addition, from the wave function
[Eq.(\ref{eq25})] and the spin currents
[Eq.(\ref{eq27},\ref{eq28})], the spin motion in the ring can be
obtained (see the discussion in Appendix).

\subsection{three conserved quantities in a ring device}

From the results of Eq.(\ref{eq27},\ref{eq28}), three quantities
characterizing the spin current ${\bf I}^n_S$ are found to be
conserved, although the spin current itself is not conserved in
the presence of a SOI. (a) The spin current ${\bf I}^n_{Sz}$ with
spin polarization along z-direction is a conserved quantity for
the ring geometry. (b) The magnitude of the spin current
$I^n_S=\sqrt{(I^n_{Sx})^2+(I^n_{Sy})^2+(I^n_{Sz})^2}
=E_aF(\theta)$ is a constant of motion. (c) For a given
cross-section of the ring, the vector of spin polarization for the
spin current ${\bf I}^n_S$ makes a fixed angle with the normal
direction of that cross-section. This angle is a constant for any
cross-section of the ring. In this sense, the spin current ${\bf
I}^n_S$ is "conserved", although the direction of spin
polarization for ${\bf I}^n_S$ is not a constant of motion due to
a SOI. So the non-conservation of spin current in the ring device
means that while moving along the ring the direction of spin
polarization is precessing due to the torque from the SOI. For the
hybrid ring, our numerical results also show that the magnitude of
spin current $I^n_S$ is again a constant of motion across the
hybrid ring, but the element of the z-direction ${\bf I}^n_{Sz}$
and the angle in (c) are not.

\section{the induced electric field by a persistent spin current}
\label{sec8}

There are a number of experiments that have been carried out to
confirm the existence of spin current,\cite{aref19,aref20,aref21}
e.g. to observe the spin current induced spin accumulations by the
Kerr effect,\cite{aref19} or to make the electric measurement
through the reciprocal spin Hall effect.\cite{aref21} Since the
persistent spin current is an equilibrium property, the above
mentioned methods are not suitable. There is also a proposal that a
spin current may cause a spin torque that can be measured
experimentally\cite{aref53,add11}. Very recently, Sonin pointed out
that this method can be employed to detect the persistent spin
current.\cite{add11,add12} On the other hand, we note that the
persistent charge current can be detected by measuring its induced
magnetic field.\cite{aref29} It has been shown that the persistent
spin current can also generate an electric
field.\cite{aref24,aref32,aref54,aref55} So this offers another way
to detect the persistent spin current by measuring its induced
electric field. In the following, we calculate the persistent spin
current induced electric field and electric potential, and show that
this electric field or the electric potential can be observed in the
present technology.

The induced electric fields $\vec{E}_{S}({\bf r})$ and
$\vec{E}_{\omega}({\bf r})$ at space point ${\bf r}=(x,y,z)$ by the
linear and angular spin currents ${\bf I}_S$ and ${\bf I}_{\omega}$
in the ring device are:\cite{aref24}
\begin{eqnarray}
 \vec{E}_S &= &-\frac{\mu_0 g \mu_B }{h} \nabla \times \int_0^{2\pi} {\bf
 I}_S(\varphi)
 \bullet \frac{{\bf r}-{\bf r}'(\varphi)}
  {|{\bf r}-{\bf r}'(\varphi)|^3} a d\varphi \label{eq29}\\
 \vec{E}_{\omega} & = &-\frac{\mu_0 g \mu_B }{h}  \int_0^{2\pi} {\bf
 I}_{\omega}(\varphi)
 \times \frac{{\bf r}-{\bf r}'(\varphi)}
  {|{\bf r}-{\bf r}'(\varphi)|^3} a d\varphi
  \label{eq30}
\end{eqnarray}
where $\mu_B$ is the Bohr magneton and ${\bf
r}'(\varphi)=a(\cos\varphi,\sin\varphi,0)$ is the position vector in
the ring. Considering the whole ring having a constant SOI, the
persistent linear and angular spin currents have been solved in the
section VII.C [see the Eqs.(\ref{eq27},\ref{eq28})]. Substituting
them into the above formulas, Eqs.(\ref{eq29},\ref{eq30}) (note that
$I_{Si}$ is the element $ I_{S\varphi i}$ of the linear spin
current) and with the help of the first and second kind complete
elliptic integral functions $K(x)$ and $E(x)$:
\begin{eqnarray}
 K(x)& =& \int_0^{\pi/2} 1/\sqrt{1-x(\sin\varphi)^2} d\varphi , \\
 E(x)& =& \int_0^{\pi/2} \sqrt{1-x(\sin\varphi)^2} d\varphi ,
\end{eqnarray}
the induced electric fields $\vec{E}_{S}({\bf r})$ and
$\vec{E}_{\omega}({\bf r})$ can be obtained straightforwardly. The
electric fields $\vec{E}_{S}({\bf r})$ and $\vec{E}_{\omega}({\bf
r})$ are rotational invariant about the $z$ axis, and in the plane
$\vec{e}_z$-$\vec{e}_r$, i.e. the elements $\vec{E}_{S\varphi}$
and $\vec{E}_{\omega\varphi}$ are zero. So here we only show
$\vec{E}_{S}({\bf r})$ and $\vec{E}_{\omega}({\bf r})$ in the
$x$-$z$ plane with ${\bf r}=(x,y,z)=(x,0,z)$:
\begin{eqnarray}
 && \hspace{-20mm} E_{Sx}  =  \frac{2c}{a x R_-^3 R_+^4}
    \left\{ K(A) R_+^2 \left[ a\left(R_+^2R_-^2 -z^2(a^2+x^2+z^2)
    \right) \cos\theta  -
  z\left(R_+^2R_-^2 + a^2(a^2-x^2+z^2)  \right) \sin\theta
  \right]\right.
  \nonumber\\
 && \left. \hspace{-15mm}
  -E(A) \left[ a\left( R_+^2R_-^2(a^2+x^2)  -16a^2x^2z^2
     \right) \cos\theta  -
  z\left(R_+^2R_-^2(2a^2+x^2+z^2) + 8a^2x^2(a^2-x^2 -z^2)  \right) \sin\theta \right]
 \right\} \\
  && \hspace{-20mm}E_{Sz}  =  \frac{2c}{a  R_-^3 R_+^4}
    \left\{ K(A) R_+^2 \left[ az(-a^2+x^2+z^2)
     \cos\theta  +
  (R_+^2R_-^2 + 2a^2 z^2) \sin\theta
  \right]\right.
  \nonumber\\
  &&  \left. \hspace{-15mm}
 + E(A) \left[ az \left( -R_+^2R_-^2  +8a^2(a^2-x^2+z^2)
     \right) \cos\theta  +
  \left(R_+^2R_-^2(3a^2-x^2-z^2) - 8a^2z^2(a^2+x^2 +z^2)  \right) \sin\theta \right]
 \right\}
\end{eqnarray}
\begin{eqnarray}
 E_{\omega x} & =& \frac{-2cz\sin\theta}{ax R_-R_+^2} \left[
    (a^2+x^2+z^2) E(A) -R_+^2 K(A) \right] \\
 E_{\omega z} & =& \frac{-2c\sin\theta}{a R_-R_+^2} \left[
    (a^2-x^2-z^2) E(A) + R_+^2 K(A) \right]
\end{eqnarray}
and $E_{Sy}=E_{\omega y}=0$. Here $R^2_{\pm} =(a\pm x)^2+z^2$, $A
=-4ax/R_-^2$, and $c=-\mu_0 g \mu_B E_a F(\theta)/h$. Then the
total electric field ${\bf E}_T ={\bf E}_S +{\bf E}_{\omega}$ are
also easily obtained, and ${\bf E}_T$ can be expressed as a
gradient of a potential $V({\bf r})$, ${\bf E}_T({\bf r}) =-\nabla
V({\bf r})$, where
\begin{equation}
 V({\bf r}) = \frac{2c}{R_-R^2_+} \left[ R_+^2 K(A) \cos\theta
  + \left((a^2-x^2-z^2)\cos\theta -2az\sin\theta \right) E(A)
  \right].
\end{equation}
In fact, this total electric field ${\bf E}_T$ can also be expressed
as:
\begin{equation}
 {\bf E}_T = -\nabla V = - c\nabla \int \vec{P}_e(\varphi) \bullet
 \frac{{\bf  r} -{\bf r}'(\varphi)}{|{\bf  r} -{\bf r}'(\varphi)|^3}
 a d\varphi ,
\end{equation}
i.e. ${\bf E}_T$ is equivalent to the electric field generated by a
1D electric dipole moment $\vec{P}_e(\varphi) =-(\cos\theta
\cos\varphi, \cos\theta\sin\varphi, \sin\theta) =-
\vec{e}_{\varphi}\times (-\sin\theta \vec{e}_r +\cos\theta
\vec{e}_z) \propto \vec{v}\times \vec{S}$ in the ring (see Fig.17c).

Fig.17 shows the electric-field lines of ${\bf E}_{S}$,  ${\bf
E}_{\omega}$, and $ {\bf E}_T$ in the $x$-$z$ plane. The across
points of the ring and the $x$-$z$ plane are at $(a,0,0)$ and
$(-a,0,0)$. The electric-field lines have the characteristics: The
field lines are in the $x$-$z$ plane and $E_{Sy}=E_{\omega
y}=E_{Ty}=0$. $\nabla \times {\bf E}_S $ and $\nabla \times {\bf
E}_{\omega}$ usually are non-zero, but the total electric fields
${\bf E}_T$ has the behavior $\nabla\times {\bf E}_T =0$, i.e.
$\oint {\bf E}_T \bullet d {\bf l} =0$. The electric fields in the
$x$-$z$ plane are mirror symmetry around $z$ axis with $E_{S/\omega/
T, x}(x,z) = -E_{S/\omega/ T, x}(-x,z)$ and $E_{S/\omega/ T, z}(x,z)
= E_{S/\omega/ T, z}(-x,z)$.

Fig.18 shows the electric-field strengths $E_{Sx/z}$, $E_{\omega
x/z}$, and $E_{Tx/z}$, along the two horizontal dashed lines [from
the point $(0,0,a)$ to the point $(4a,0,a)$, or from the point
$(0,0,0.1a)$ to the point $(2a,0,0.1a)$] in the Fig.17. In this
calculation, we consider that only the lowest level $n=0$ (the
ground state) in the ring device is occupied by the electron, i.e.
taking the parameter $E_0 < E_F <E_1$, and in this case ${\bf I}_S=
{\bf I}_S^0$ and ${\bf I}_{\omega} ={\bf I}_{\omega}^0$. We also
take the parameters g factor $g=2$ and the efficient electron mass
$m=0.036m_e$. At the point $x=0$, the $x$-direction element
$E_{S/\omega/ T, x}=0$ due to the mirror symmetry, but the
$z$-direction element $E_{S/\omega/ T, z}$ keep quite large value
still. In slightly far away from the ring device (e.g. $z=a$ in
Fig.18a, c, e), the electric field ${\bf E}_{\omega}$ induced from
the persistent angular spin current is in the same order with the
electric field ${\bf E}_{S}$ induced from the persistent linear spin
current. So ${\bf E}_{\omega}$ is important for contributing to the
total electric field ${\bf E}_T$. On the other hand, while very near
the ring device (e.g. $z=0.1a$ in Fig.18b, d, f), ${\bf E}_{S}$ is
much larger than ${\bf E}_{\omega}$ and ${\bf E}_{S}$ is dominant in
${\bf E}_{T}$. It is worth to mention that the total electric field
${\bf E}_T$ can reach $10^{-2} V/m$ at the point $(a,0,0.1a)$ which
is $0.1a=5nm$ over the ring $(a,0,0)$ (see Fig.18f). Also let us
estimate the electric potential difference due to ${\bf E}_T$, this
potential difference between two points $(a,0,0.01a)$ and
$(a,0,0.01a)$ is about $1nV$. Although this potential value is very
small, it is measurable in the present
technology.\cite{aref24,aref32}

\section{the spin-spin interaction and the conserved persistent spin current}
\label{sec9}

In previous sections single spin picture is adapted, i.e., there is
no spin-spin interaction, we find that in general the spin current
is not conserved. In this section, we demonstrate that if one
includes strong spin-spin interaction, the spin current will be
conserved. In this case, if a spin precesses, the response of other
spins will be against it, i.e, they will precesses in opposite
directions. As a result the total spin precession is zero everywhere
and therefore the spin current is automatically conserved by using
the conventional definition.\cite{aref27,add13} To account for the
spin-spin interaction, the Hamiltonian $H$ is:
\begin{equation}
  H = \sum\limits_i H_0({\bf r}_i) + \sum\limits_{i,j} J({\bf r}_i,{\bf
  r}_j) {\bf S}({\bf r}_i) \bullet {\bf S}({\bf r}_j),
  \label{eq39}
\end{equation}
where $H_0({\bf r})$ is the one-body Hamiltonian (e.g. the
Hamiltonian in Eq.(1) for the ring device) and the second term is
the spin-spin interaction. Usually, it is very difficult to solve
this Hamiltonian because of the many-body interaction,
$\sum\limits_{i,j} J({\bf r}_i,{\bf r}_j) {\bf S}({\bf r}_i) \bullet
{\bf S}({\bf r}_j)$. By introducing an induced self-consistent field
${\vec H}_1$ that could be due to spin-spin
interaction,\cite{aref27} the many-body Hamiltonian $H$ in
Eq.(\ref{eq39}) reduces into the one-body form:
\begin{equation}
  H({\bf r}) = H_0({\bf r}) + \hat{\sigma}\bullet \vec{H}_1({\bf r})
  \label{eq40}
\end{equation}
Now the Hamiltonian of Eq.(\ref{eq40}) is easily solved, and the
linear spin current and the spin torque all depend on ${\vec H}_1$.
The induced self-consistent field ${\vec H}_1$ in Eq.(\ref{eq40}) is
determined by requiring that the spin torque (or the angular spin
current) is zero or ${\bf I}_w({\vec H}_1)=0$ for any ${\bf r}$.
Once ${\vec H}_1$ is solved the persistent spin current is
automatically conserved by using the conventional definition.

Now we apply this method to the ring without the normal region and
set SOI to a constant with $\alpha_R(\varphi)=\alpha_R$. We start
with the following Hamiltonian
\begin{equation}
H=H_0+\sigma_r H_1
\end{equation}
where $H_0$ is the original Hamiltonian [see Eq.(1)] for a ring with
full SOI. The reason that we choose the induced field as $\sigma_r
H_1$ is because we know $I_z$ is conserved and there is a torque
along ${\hat e}_\varphi$. So a self-consistent induced magnetic
field is needed (a term $\sigma_r H_1$) to balance the torque. The
eigenfunction of the new Hamiltonian is the same as Eq.(\ref{eq25})
given by
\begin{equation}
 \Psi_n(\varphi) = \left(\begin{array}{l}
       \cos (\theta/2) e^{i n\varphi} \\
       -\sin (\theta/2) e^{i(n+1)\varphi}
       \end{array}
       \right),
\end{equation}
where the eigenvalue looks the same as before
\begin{equation}
E_n = E_a [n^2 + (n+1/2) (1-1/\cos\theta)]
\end{equation}
but $\theta$ takes a different value,
\begin{equation}
\tan\theta = \frac{\alpha}{aE_a} +\frac{H_1}{E_a(n+1/2)}
\label{rel1}
\end{equation}
Note that if the self-consistent field $H_1$ is zero, $\tan\theta$
recovers the non-interacting case. Using the conventional definition
the persistent spin current with polarization in three directions
are easily calculated and found to be
\begin{eqnarray}
I_{Sx} &=& -E_a F_1(\theta) \sin\theta\cos\varphi \nonumber \\
I_{Sy} &=& -E_a F_1(\theta) \sin\theta\sin\varphi \nonumber \\
I_{Sz} &=& -E_a [n+1/2-1/(2\cos\theta)] \cos\theta
\end{eqnarray}
where
\begin{equation}
F_1(\theta)=[n+1/2-\alpha/(2aE_a\sin\theta)]/2\pi \label{fun}
\end{equation}
If $H_1=0$, $F_1(\theta)=[n+1/2-1/(2\cos\theta)]/2\pi =F(\theta)$
that is the previous result without the spin-spin torque
interaction. Now we calculate the total torque. From continuity
equation, the torque is just the angular spin current ${\bf
I}_\omega=Re[\Psi^\dagger \hat{\vec{\omega}} \times \hat{s} \Psi]$
with $\hat{\vec \omega} = H_1 \vec{e}_r - (i\alpha/a)
\partial/\partial \varphi ~ \vec{e}_\varphi
 \times \vec{e}_z$. It is easy to show that ${\bf I}_\omega$ has only ${\vec e}_\varphi$
component
\begin{equation}
I_{\omega \varphi} = H_1 \cos\theta +\frac{\alpha}{a} (n \cos\theta
- \sin^2\theta/2)
\end{equation}
Note that the purpose of introducing the self-consistent field $H_1$
is to make sure that the spin current is conserved or the torque
vanishes as we have discussed earlier. Setting the torque $I_{\omega
\varphi}$ to zero we obtain the second equation that determines the
self-consistent field $H_1$
\begin{eqnarray}
H_1 &=& -\frac{\alpha}{a\cos\theta} (n \cos\theta -\sin^2\theta/2) \nonumber \\
&=& -\frac{\alpha}{a} (n+1/2-1/(2\cos\theta)) \label{rel2}
\end{eqnarray}
Once $H_1$ is solved from Eqs.(\ref{rel1}) and (\ref{rel2}), the
spin current will be conserved following the continuity equation.
Plugging Eq.(\ref{rel2}) into Eq.(\ref{rel1}), we find
\begin{equation}
(n+1/2)\sin\theta = \frac{\alpha}{2aE_a} \label{rel3}
\end{equation}
This means that $F_1(\theta)$ defined in Eq.(\ref{fun}) is zero and
the conserved persistent spin current is nonzero only for the spin
polarization along z-direction
\begin{eqnarray}
I_{Sz} = -E_a [n+1/2-1/(2\cos\theta)] \cos\theta
\end{eqnarray}
where $\theta$ is determined by Eq.(\ref{rel3}).

We wish to point out that even for the hybrid ring, the above
approach can be used and the persistent spin current is also
conserved in the presence of spin-spin interaction. The only
difference is that we have to introduce three self-consistent fields
$H_i$, $i=1,2,3$. From the energy dispersion relation, we have
relationship between $\theta$ and $H_i$. By requiring the torque
along each direction to be zero, we obtain three additional
equations. These four equations will determine $H_i$ and $\theta$.
This in turn gives the displacement spin current and hence the
conserved persistent spin current.

\section{conclusion}
\label{sec10}

In summary, we have investigated two closely related subjects: (a)
the prediction of a pure persistent spin current in an equilibrium
mesoscopic device with solely spin-orbit interaction (SOI), and (b)
the issues concerning the definition of the spin current. Through
the physical arguments and physical pictures from four different
aspects, the analytic calculation results of a SOI-normal hybrid
ring, as well as the discussion of the sharp interface between the
normal and SOI part, we demonstrated that the persistent spin
current indeed exists in the equilibrium device with a SOI alone. In
particularly, we emphasize that this persistent spin current is an
analog of the persistent charge current in the mesoscopic ring
threaded by a magnetic flux, and it describes the real spin motion
and are experimentally measurable.

After showing the existence of the persistent spin current, we
investigate the definition of the spin current. We point that: (i)
the non-zero spin current in the equilibrium SOI's device is the
persistent spin current; (ii) in general the spin current is not
conserved; and (iii) the Onsager relation is violated for the spin
transport, and in particularly, it can not be recovered through
modification of the definition of the spin current. So these three
"flaws", the non-zero spin current in the equilibrium case, the
non-conserved spin current, and the violation of the Onsager
relation, of the conventional definition of the spin current are
intrinsic properties of spin transport. In particular, the
conventional definition, ${\bf I}_S = Re \Psi^{\dagger} \hat{v}
\hat{s} \Psi$, possesses a very clear physical picture, and is
capable of describing the spin motion. So we draw the conclusion
that the conventional definition of the spin current makes
physical sense, and no need to modify it.

In addition, a number of problems have also been discussed. The
relation between the persistent spin current and transport spin
current is discussed, and we find that they are indistinguishable in
the coherent part of the device. We calculate the persistent linear
and angular spin current in the SOI's region of the hybrid ring, and
the results show that the persistent spin current still exists in
the SOI's region, even when the SOI covers the whole ring. The
measurement issue of the persistent spin current is also discussed,
we suggest that the persistent spin current can be observed by
detecting its induced electric field. In the presence of a spin-spin
interaction in the ring, we find that the persistent spin current
using the conventional definition is automatically conserved.

\section*{Acknowledgments}

We gratefully acknowledge the financial support from NSF-China under
Grant Nos. 10474125, 10525418, and 60776060 (Q.F.S.); US-DOE under
Grant No. DE-FG02-04ER46124 and NSF under CCF-052473 (X.C.X.); a RGC
grant from the Government of HKSAR grant number HKU 7044/05P (J.W.).

\section*{Appendix}

In this appendix, we analyze the motion of the spin which relates to
the persistent spin current in the equilibrium. For simplicity, we
consider the constant SOI case, and the eigen wave-function has been
solved in Eq.(\ref{eq25}). From this wave-function, the spin
$\vec{S} =\langle\Psi_n|\hat{s}|\Psi_n\rangle
=\frac{\hbar}{2}(-\sin\theta \cos \varphi, -\sin\theta \sin \varphi,
\cos\theta)$. This spin vector is in the $\vec{e}_z$-$\vec{e}_r$
plane, and its angle with the $\vec{e}_z$ (i.e. $z$) axis is
$-\theta$. Then the spin motion can also be obtained
straightforwardly by solving the velocity and its angular velocity
$\vec{\omega}$. The direction of spin (translational) motion is
counter-clockwise, while precessing with $\vec{\omega}$ in the
perpendicular direction of $\vec{S}$, so that the spin is in the
$\vec{e}_z$-$\vec{e}_r$ plane all along. The element of the spin in
the $x$-$y$ plane and its motion are shown as in Fig.12b. On the
other hand, for the time-reversal state $T\Psi_n$, the spin
direction, its motion direction, and the precession direction, all
reversed, as shown the clockwise arrow in Fig.12b. But the spin
current of $T\Psi_n$ is completely same with that of $\Psi_n$.
Therefore the persistent spin current indeed describes the real
motion of the spin.

In fact, besides the ring geometry, the device also can be other
shape. For example, we have analyzed the spin motion and the
persistent spin current in the quasi one-dimensional equilibrium
quantum wire.\cite{aref39} Similar conclusions can be drawn.

\newpage

\begin{figure}
\caption{ (Color online) (a) and (b) are the schematic diagrams
for a mesoscopic ring with a magnetic atom or an ion at its
center. (c) Schematic diagram for a hybrid mesoscopic ring having
the Rashba SOI in part of the ring while the other part being
normal. } \label{fig:1}

\caption{ (Color online) (a), (b), (c), and (d) are schematic
diagrams for the devices of the Hall effect, the spin Hall effect,
the persistent (charge) current, and the persistent spin current,
respectively.} \label{fig:2}

\caption{ (a) and (b) show the eigen energies $E_n$ vs. $\alpha_R$
for $\Phi_0=\pi$ and vs. $\Phi_0$ for $\alpha_R=3\times 10^{-11}
eVm$, respectively. The radius of the ring $a=50nm$. }\label{fig:3}

\caption{ $I^n_{Sy}$ (a) and $I^n_{Sz}$ (b) vs $\alpha_R$ for
$\Phi_0=\pi$ and $a=50nm$. Along the arrow direction, $n=7$, $5$,
$3$, $1$, $0$, $2$, $4$, $6$, and $8$. Here the level index $n=0$,
$1$, $2$, ..., represent the ground state, the first excited state,
the second excited state, ..., respectively. } \label{fig:4}

\caption{ $I^n_{Sx}$ (a), $I^n_{Sy}$ (b), and $I^n_{Sx}$ (c) vs.
$\Phi_0$ for $\alpha_R=3\times10^{-11}eVm $ and $a=50nm$. Along the
arrow direction, $n=6$, $4$, $2$, $0$, $1$, $3$, and $5$. Here the
level index $n=0$, $1$, $2$, ..., represent the ground state, the
first excited state, the second excited state, ..., respectively. }
\label{fig:5}

\caption{ (Color online) (a) and (b) show $I_{Sy}$ and $I_{Sz}$ vs
$\alpha_R $ for $\Phi_0=\pi$. (c) and (d) show $I_{Sx/y/z}$ vs the
angle of the normal region $\Phi_0$ for $\alpha_R=3\times 10^{-11}
eVm$ and $E_f=3E_a$ (c) and $6E_a$ (d). The radius of the ring
$a=50nm$ and temperature $T=0$. } \label{fig:6}

\caption{ (Color online) (a) and (b) show $I_{Sy}$ and $I_{Sz}$ vs
$\alpha_R $ at the different temperature $T$. The radius of the ring
$a=50nm$ and $\Phi_0=\pi$. } \label{fig:7}

\caption{ (Color online) The eigen-energies $E_n$ vs. $\alpha_R$.
The red dotted curves are for the tight-binding model of the
discrete Hamiltonian with the lattice points $N=20$ (a), $N=50$ (b),
and $N=150$ (c). The black solid curves are from the exact method
used in the section III, i.e. these black solid are completely the
same as the curves in Fig.3a. All parameters are the same as in
Fig.3a. } \label{fig:8}

\caption{ (Color online) $I^n_{Sy}$ vs. $\alpha_R$. The red dotted
curves are for the tight-binding model of the discrete Hamiltonian
with the lattice points $N=20$ (a), $N=50$ (b), and $N=150$ (c). The
black solid curves are for the exact method used in the section III,
i.e. these black solid are completely the same as the curves in
Fig.4a. All parameters are the same as in Fig.4a. } \label{fig:9}

\caption{ (Color online) (a) the eigen energies $E_n$ vs.
$\alpha_R$, (b) $I^n_{Sy}$ vs. $\alpha_R$, (c) the persistent spin
current $I_{Sy}$ vs. $\alpha_R$, and (d) the persistent spin current
$I_{Sz}$ vs. $\alpha_R$ for the ring with a non-sharp interface (see
the text). The ring radius $a=50nm$ and the number of lattice points
$N=200$. In (b), along the arrow direction, $n=7$, $5$, $3$, $1$,
$0$, $2$, $4$, $6$, and $8$. In (c) and (d), the temperature $T=0$.
} \label{fig:10}

\caption{ (Color online) (a) is a schematic diagram for the ring
device threaded by a magnetic flux and coupled to left and right
leads. (b) is a schematic diagram for the ring device with the SOI
and coupled to left and right leads. } \label{fig:11}

\caption{ (Color online) (a) Schematic diagram for a spin movement
in a terminal device, in which the spin moves along the x axis, and
flips about $x=0$. (b) Schematic diagram for the spin movement and
precession in a ring. } \label{fig:12}

\caption{ (Color online) (a) and (b) are schematic diagrams for
the base vectors $\vec{e}_x$ and $\vec{e}_y$, and $\vec{e}_x$ and
$\vec{e}_r$. (c), (d), and (e) are schematic diagrams for the spin
potential $V_{sx}$, $V_{sy}$, and $V_{sz}$, respectively. }
\label{fig:13}

\caption{ (Color online) The linear spin current $I^n_{Sx/y/z}$
and the angular spin current $I^n_{\omega\varphi/z}$ vs the angle
coordinates $\varphi$ for the level $n=0$ (a and d), 1 (b and e),
and 2 (c and f). The parameters are $\alpha_R =3\times
10^{-11}eVm$, $\Phi_0 =\pi$, and the ring radius $a=50nm$. The
solid (dark) curve, dashed (blue) curve, and dotted (red) curve in
panels (a, b, and c) correspond to $I^n_{Sx}$, $I^n_{Sy}$, and
$I^n_{Sz}$, respectively. The solid (dark) curve and dotted (red)
curve in panels (d, f, and e) correspond to $I^n_{\omega\varphi}$
and $I^n_{\omega z}$, respectively. } \label{fig:14}

\caption{ (Color online) The persistent spin currents $I_{Sx/y/z}$
vs. the angle coordinates $\varphi$ for the Fermi energy $E_F =2E_a$
(a) and $6E_a$ (b). The temperature $T=0$ and the other parameters
are same with Fig.14. The solid (dark) curve, dashed-dotted (blue)
curve, and dotted (red) curve correspond to $I_{Sx}$, $I_{Sy}$, and
$I_{Sz}$, respectively.
 } \label{fig:15}

\caption{ (Color online) The persistent spin currents $I_{Sx/y/z}$
vs. the angle coordinates $\varphi$ for $\Phi_0=\pi/2$ (a), $\pi/4$
(b), and $0$ (c). The temperature $T=0$ and the other parameters are
same with Fig.14. The solid (dark) curve, dashed-dotted (blue)
curve, and dotted (red) curve correspond to $I_{Sx}$, $I_{Sy}$, and
$I_{Sz}$, respectively.
 } \label{fig:16}

\caption{ (Color online) Schematic plots of electric-field lines
of the electric fields $\vec{E}_{\omega}$ (a), $\vec{E}_{S}$ (b),
and $\vec{E}_{T}$ (c) for the case when the entire ring has a
constant $\alpha_R$. } \label{fig:17}

\caption{ (Color online) The electric field strengthes $E_{\omega
x/z}$ (a and b), $E_{S x/z}$ (c and d), and $E_{T x/z}$ (e and f)
vs. the position $x$ along the horizontal dashed lines in Fig.17
with $z=a$ (a, c, e) and $z=0.1a$ (b, d, f). The parameters are
$\alpha_R = 3\times 10^{-11} eVm$, $\Phi_0 =0$, the ring radius
$a=50nm$, and only the lowest level $n=0$ occupied. The solid
curve and the dotted curve correspond to $E_{\omega/S/T, x}$ and
$E_{\omega/S/T, z}$, respectively. } \label{fig:18}
\end{figure}


\begin{references}
\bibitem{note1}
Note that this conclusion is not in contradiction with the
opinions in our previous papers, Ref.[\cite{aref24,aref27}]. The
detail discussion can be found in the section VI.
\bibitem{note3}
Consider the case that there is a point charge at the center of
the ring that gives rise to the Thomas SOI. We find that the
persistent spin current also exists, similar as for the Rashba SOI
case discussed in the text.
\bibitem{note4}
In fact, the boundary condition of the flux can be deduced from
the Sch$\ddot{o}$dinger equation, by taking the integral
$\int^{0^+}_{0_-} H(\varphi) \Psi(\varphi) d\varphi =
\int^{0^+}_{0_-} E \Psi(\varphi) d\varphi $.
\bibitem{note5}
Note that the eigen level of $n=0$ is two-fold degenerate, and
this level does not split for non-zero $\alpha_R$.
\bibitem{note6}
For a continuous wave-function $\Psi(\varphi)$, its normalization
equation is $\int_0^{2\pi} \Psi^{\dagger}(\varphi) \Psi(\varphi)
ad\varphi =1$. For a discrete wave-function $\Psi_j$, its
normalization equation is $\sum\limits_j \Psi_j^{\dagger} \Psi_j
a\Delta\varphi =1$.
\bibitem{note7}
Because of the two-fold degeneracy for each level $E_n$, we have
multiplied the factors 2 in the plotting of $I^n_{Si}$, i.e.
$I^n_{Si} =2 Re \Psi^{\dagger}_n \hat{v}\hat{\sigma}_i \Psi_n$ and
$I_{Si} =\sum\limits_n I^n_{Si}f(E_n)$.
\bibitem{note8}
Since the size of the ring is assumed to be within the coherent
length, the transport charge and spin currents are also without
dissipation in the ring.
\bibitem{note9}
By the way, we emphasize that the implications of the angular spin
current and the spin torque are different, although they are
completely same in terms of numerical value. The angular spin
current describes the spin motion or the spin flow, but the spin
torque expresses the action on the spin. This is similar to the
relation between the force $\vec{F}$ and the acceleration
$\vec{a}$. If taking the electron mass $m=1$ as the unit, the
value of the force equals that of the acceleration, but their
physical meanings are different.
\bibitem{add13}
When the total spin precession is everywhere zero, the angular spin
current ${\bf I}_{\omega}=0$, and the spin continuity equation
becomes: $\nabla \bullet {\bf I}_{Si} +d S_i/dt =0$, where $S_i$ is
the spin accumulation and is zero in the present equilibrium case.
Note that this continuity equation for every spin element $i=x,y,z$
is completely the same as the charge continuity equation. Therefore,
the conserved spin current can be derived from Noether's theorem,
similar as the conserved charge current.
\end{references}
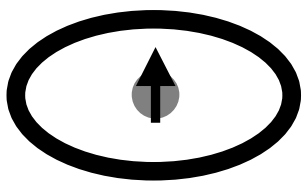
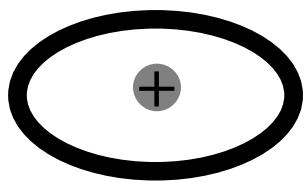
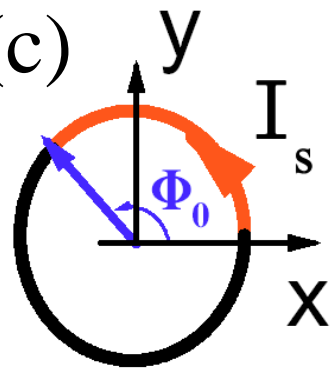
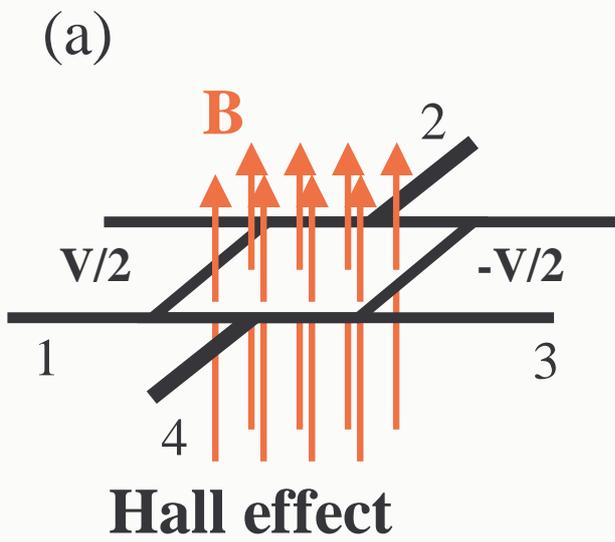
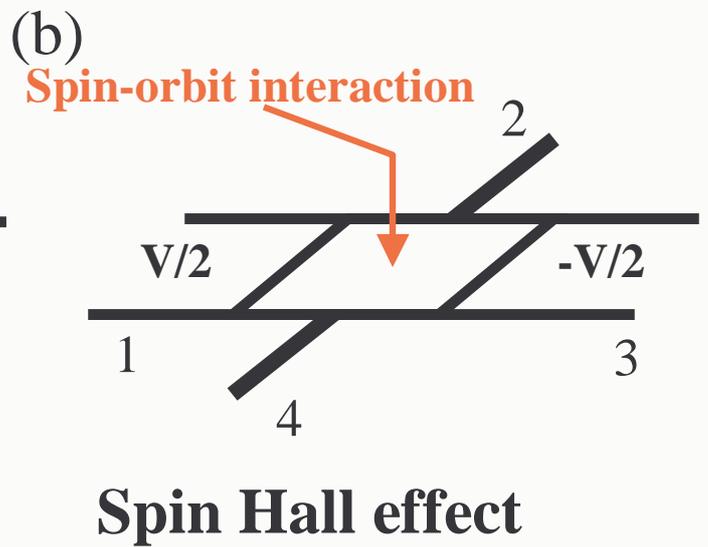
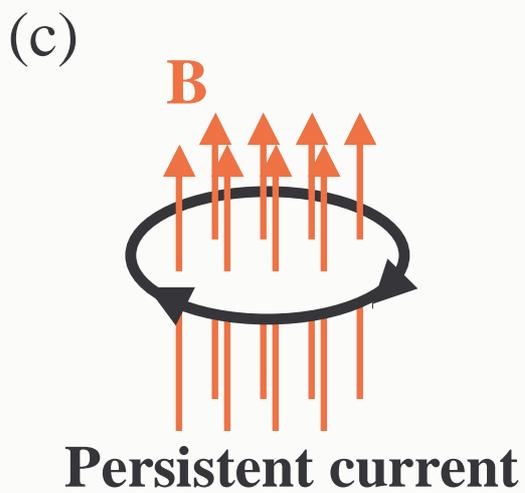
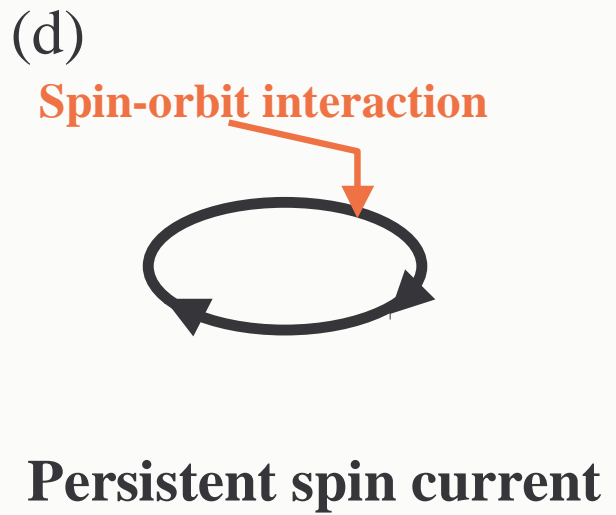
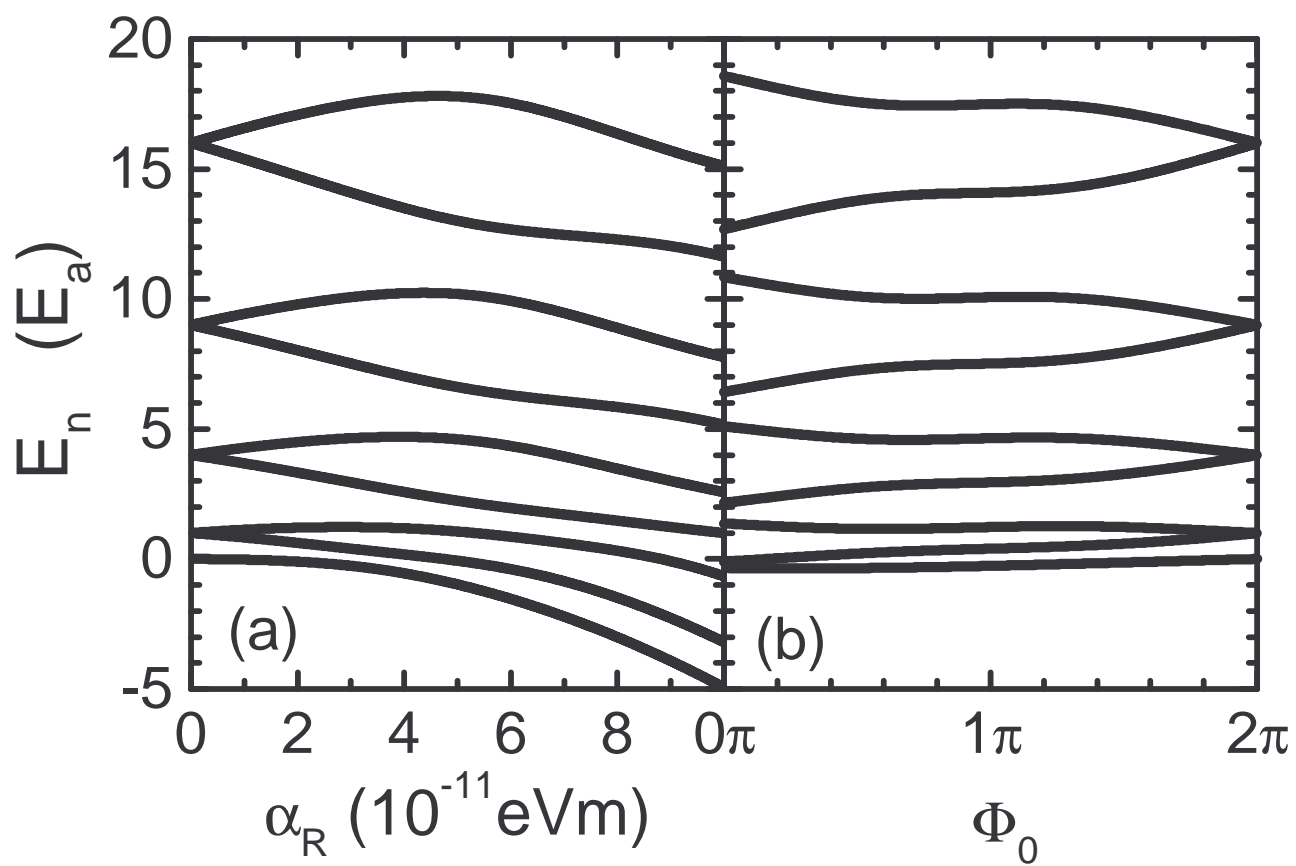
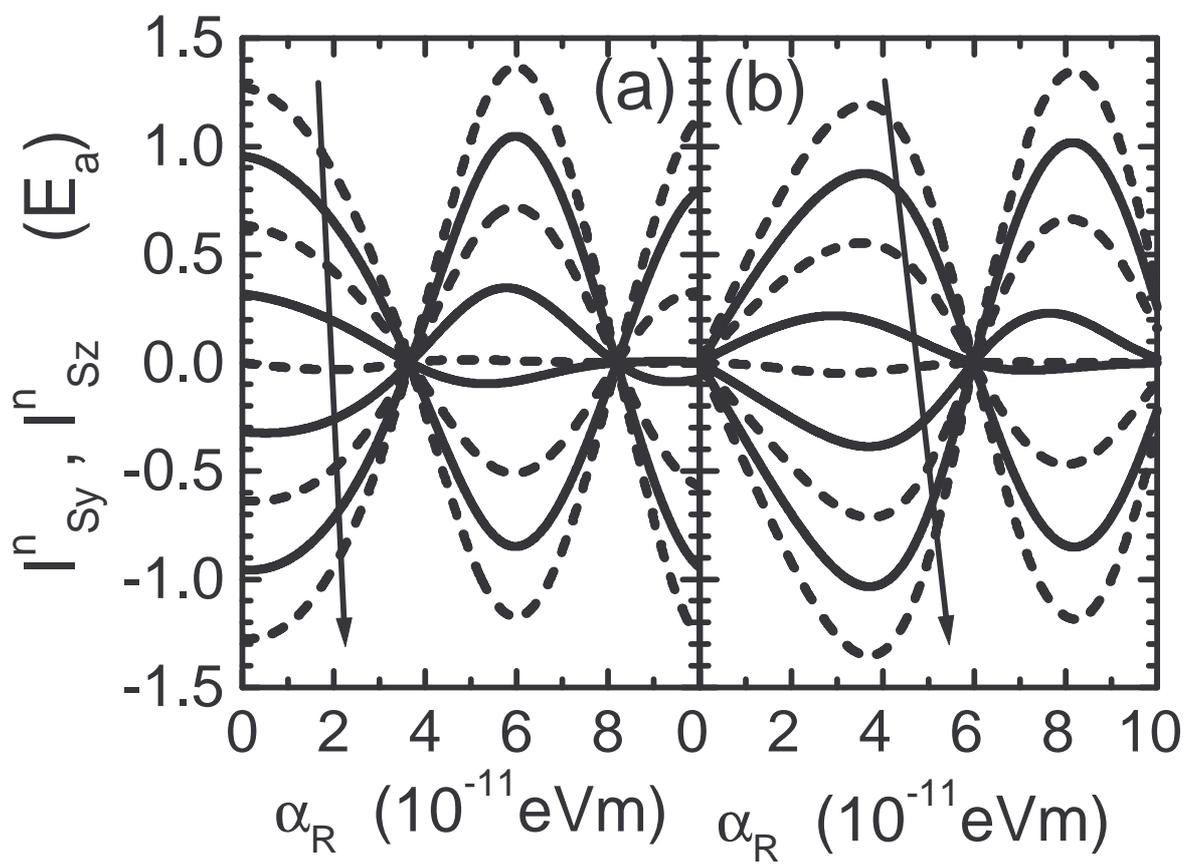
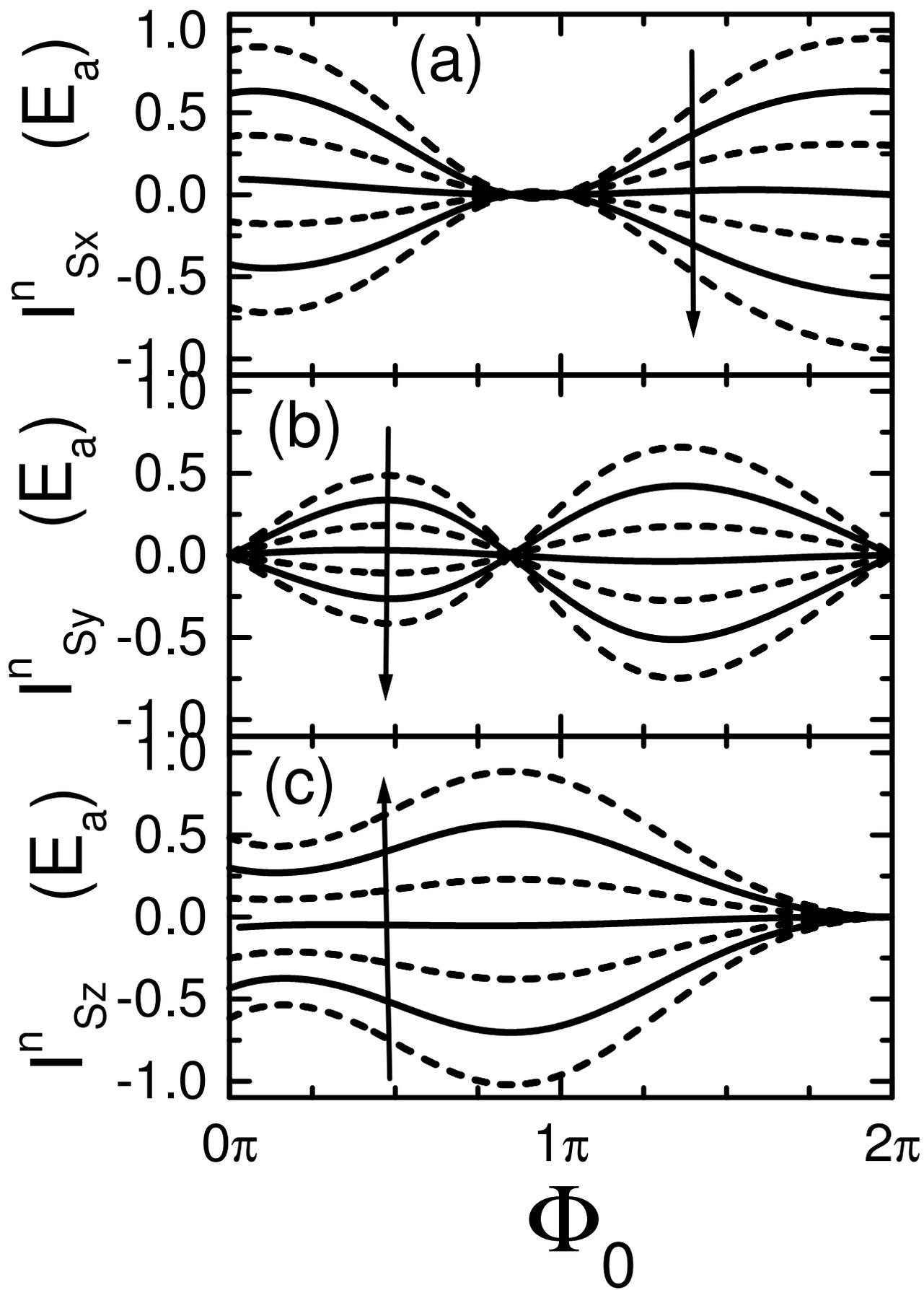
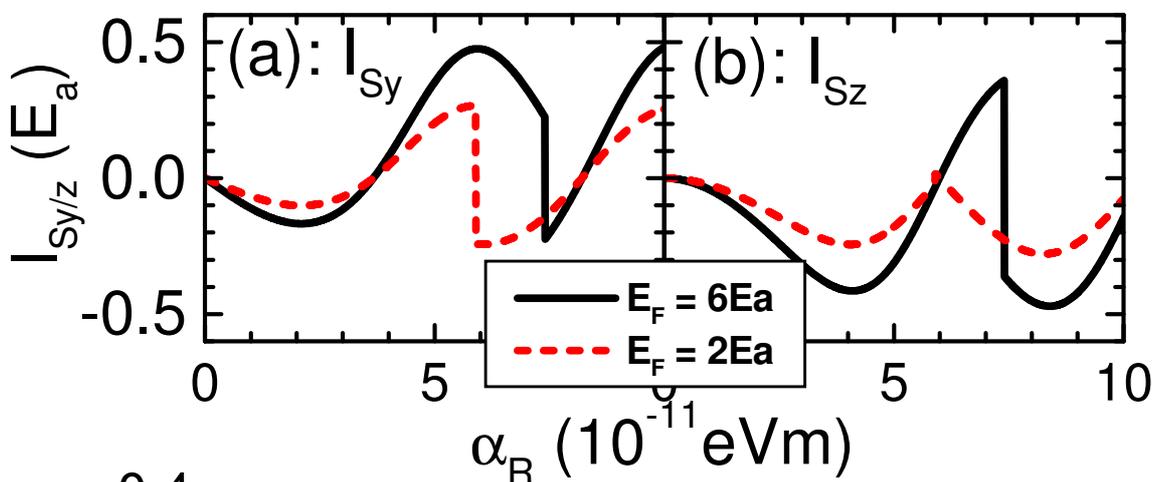
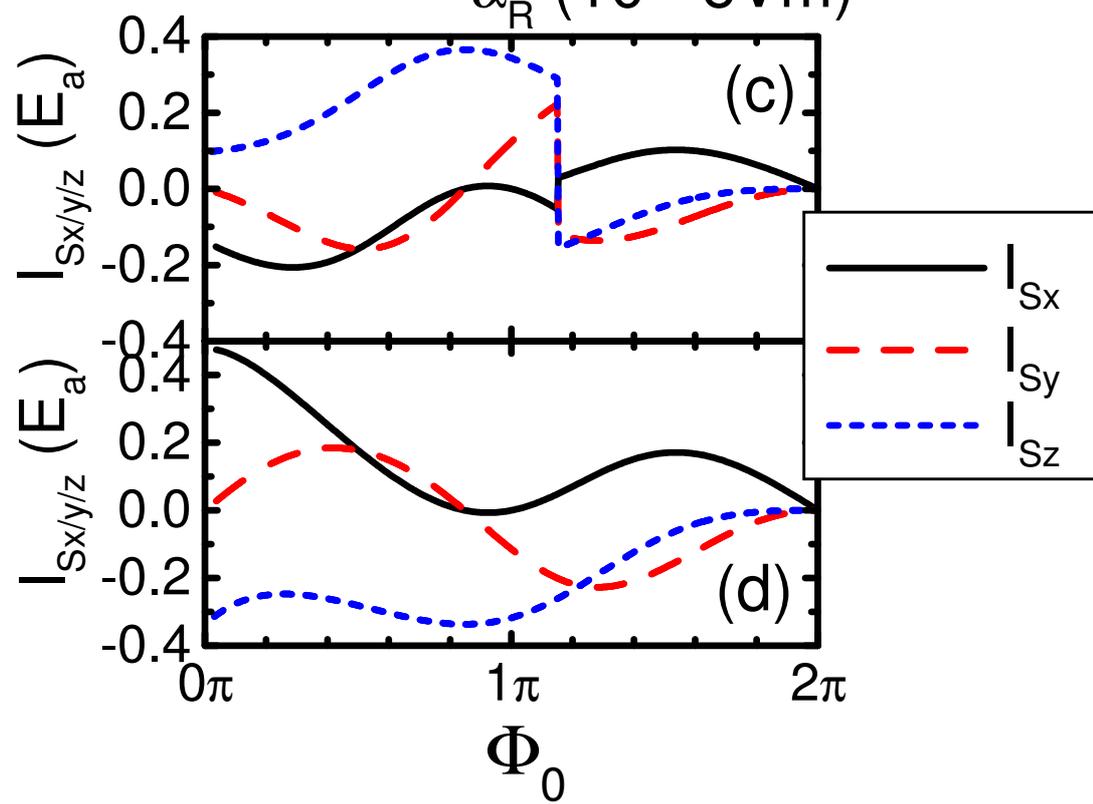
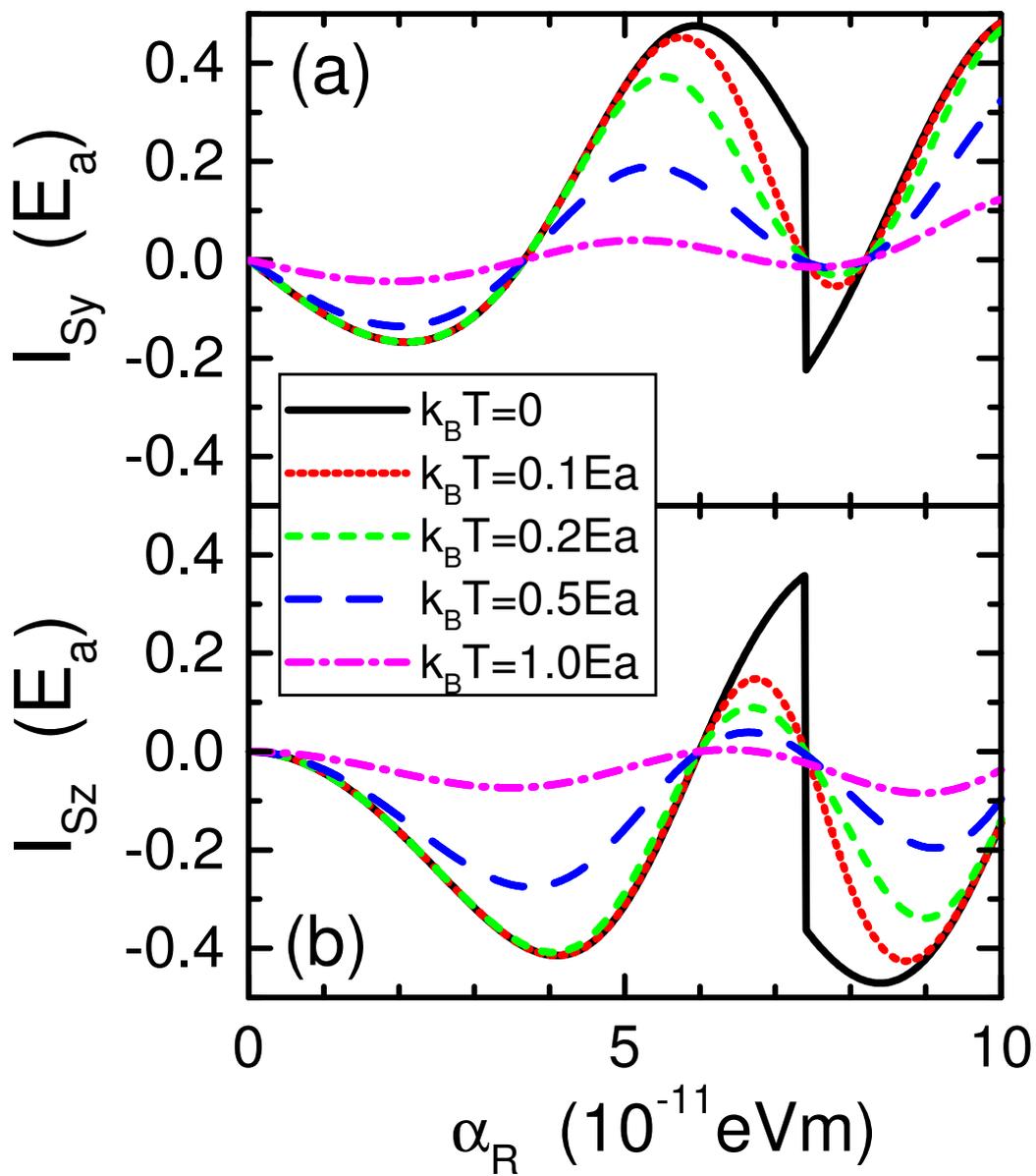
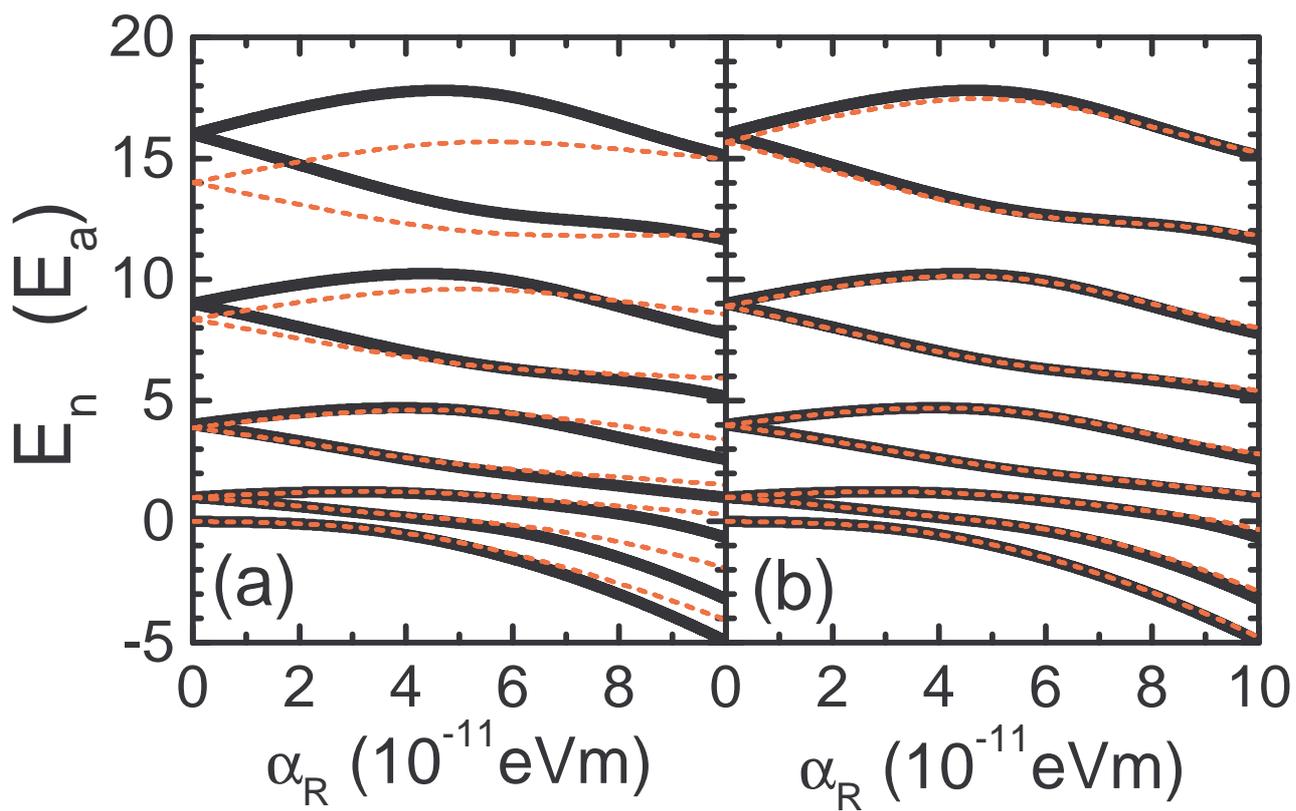
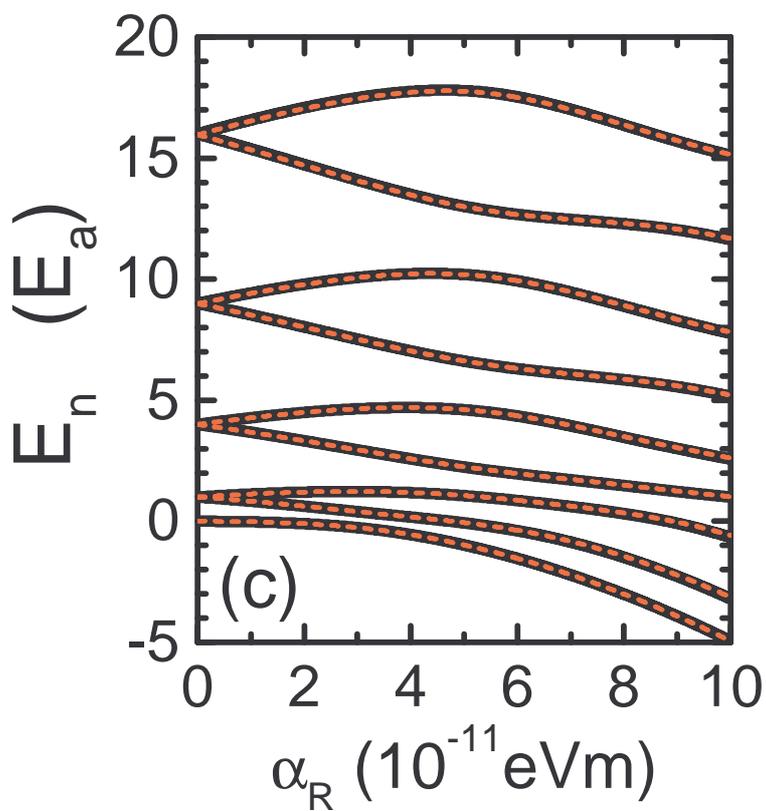
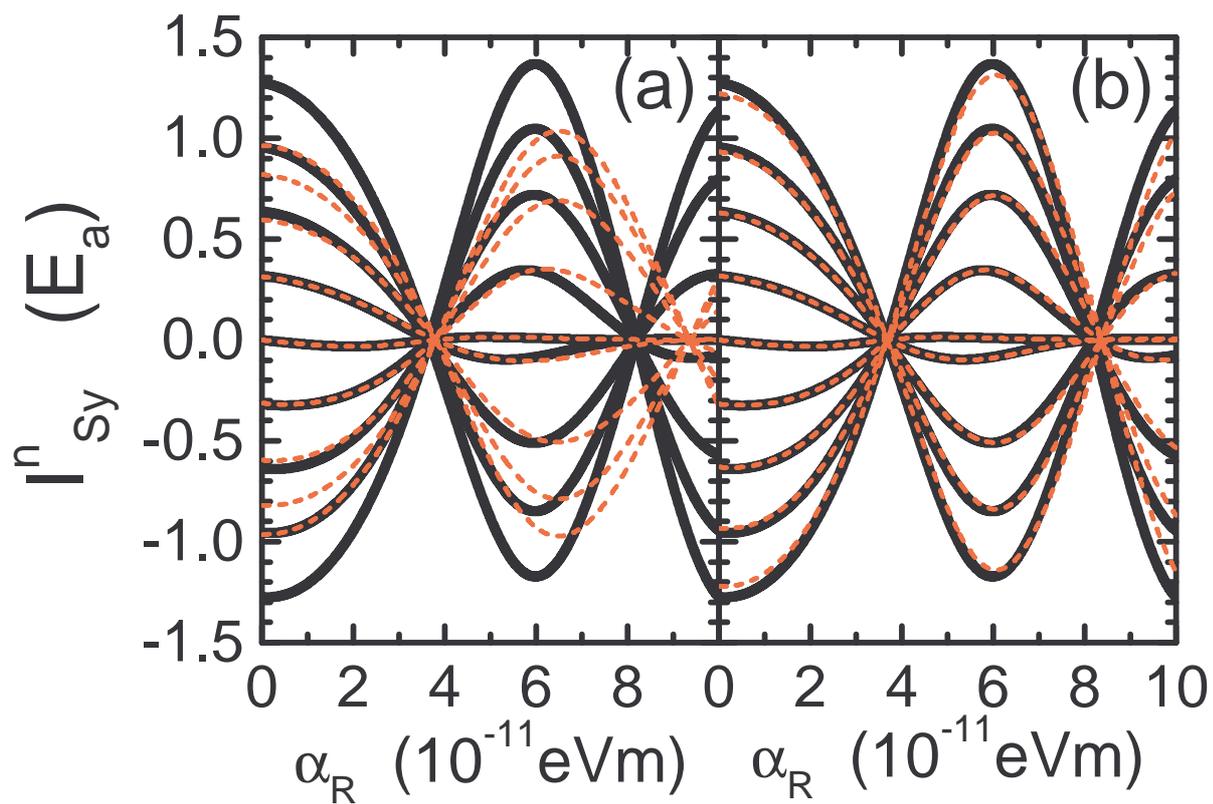
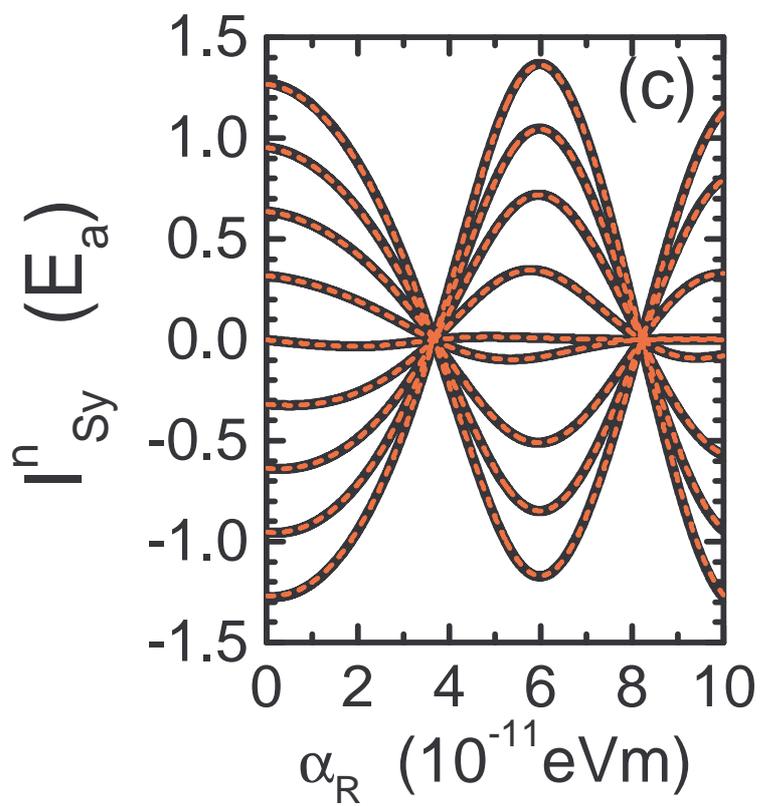
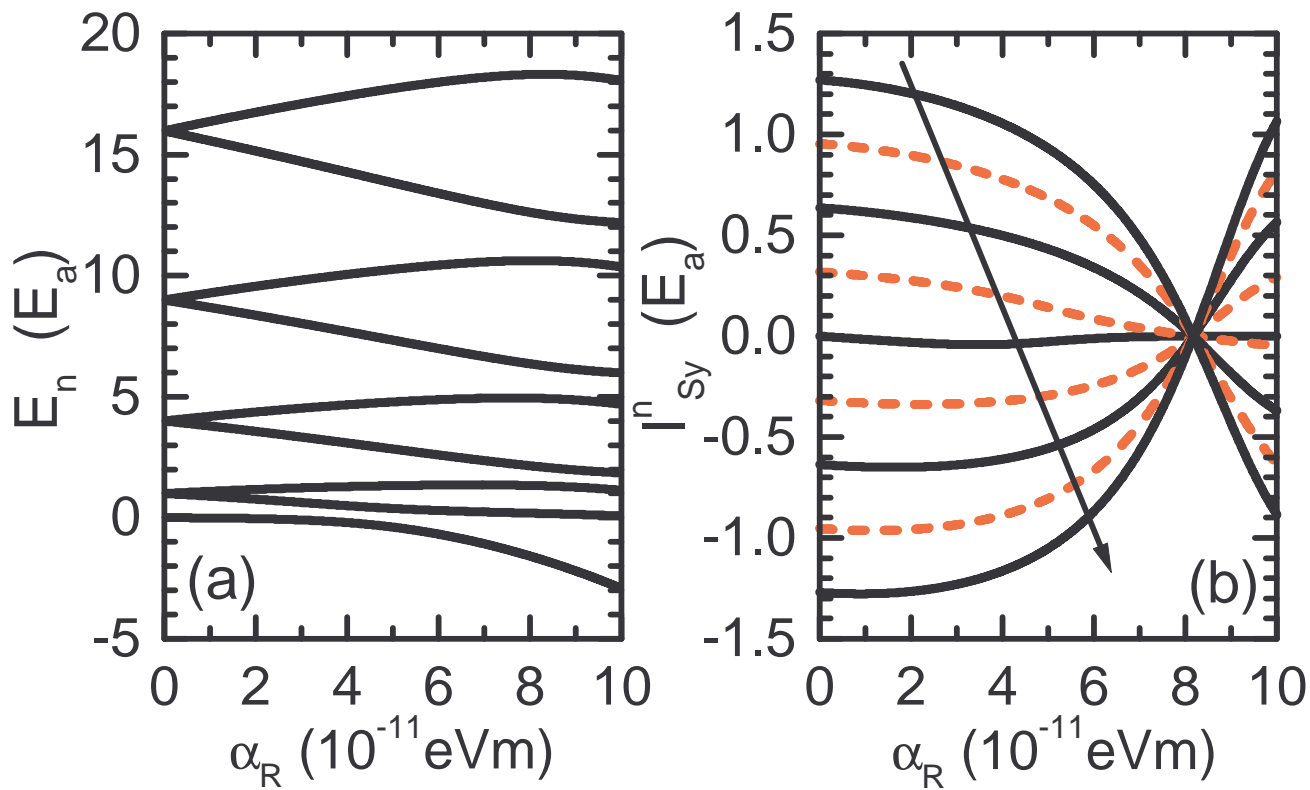
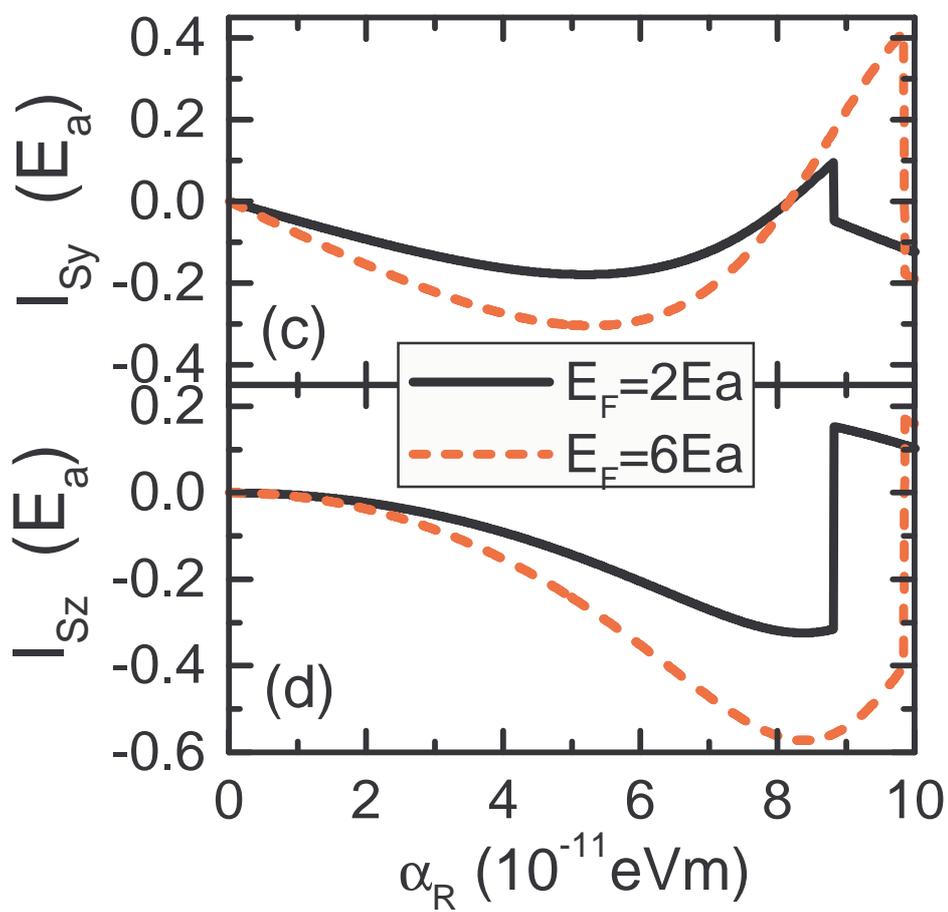
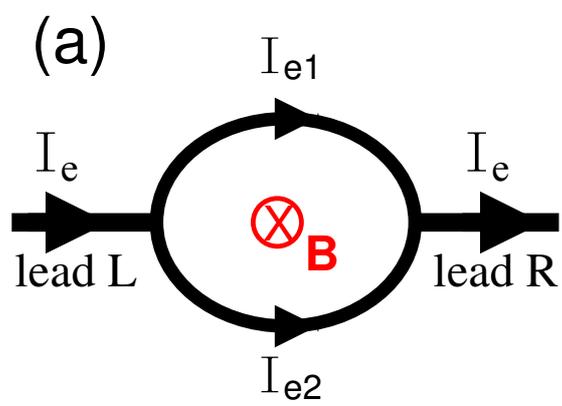
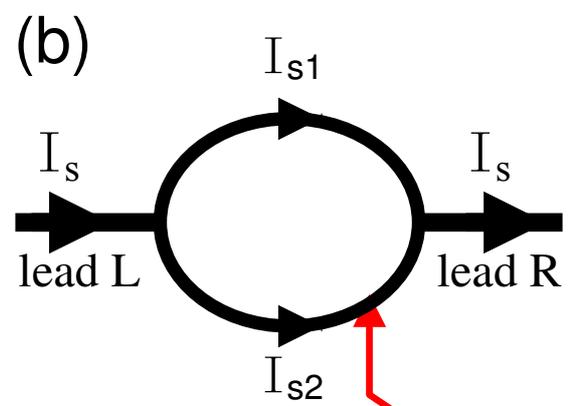
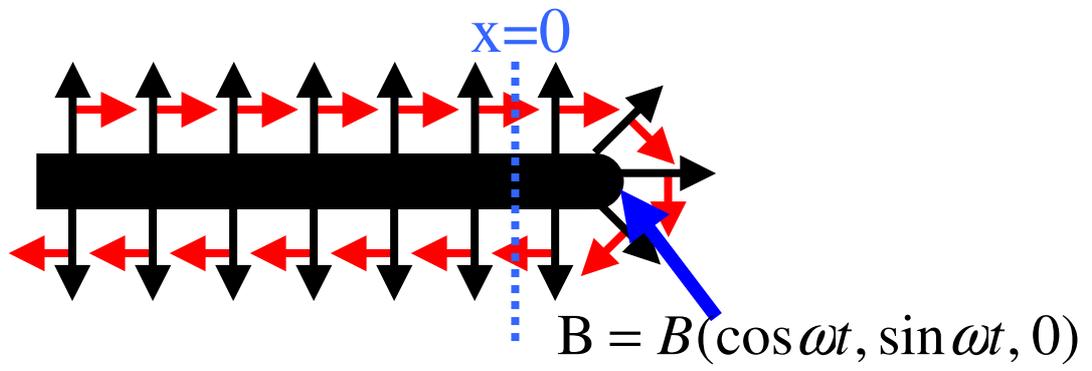
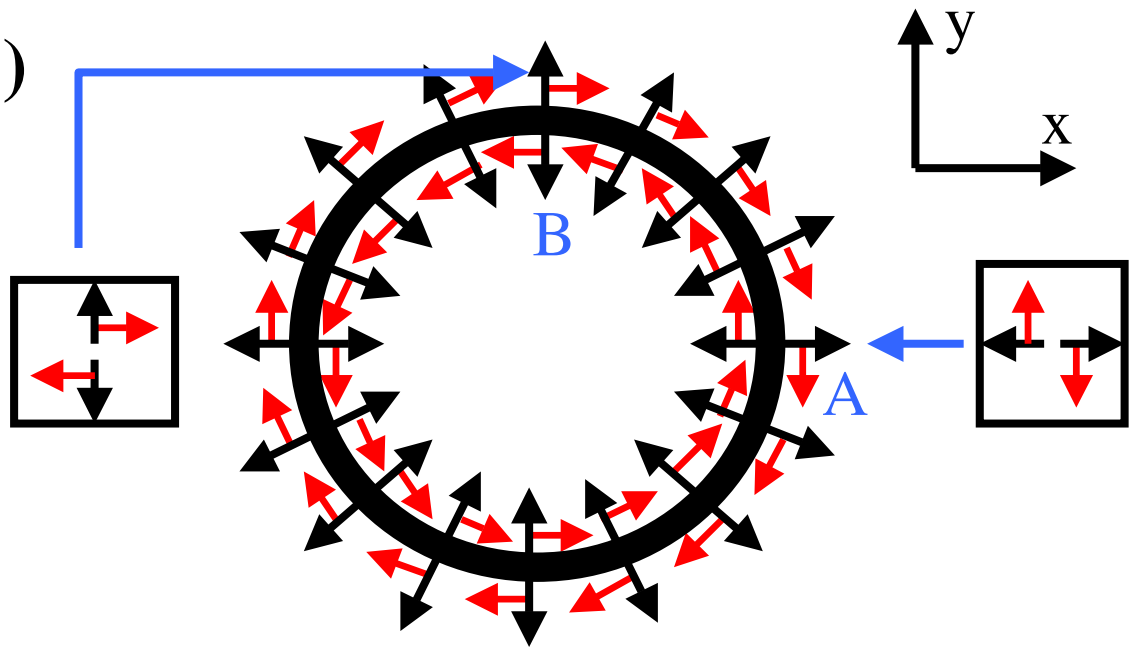
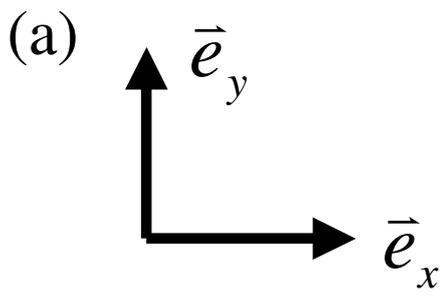
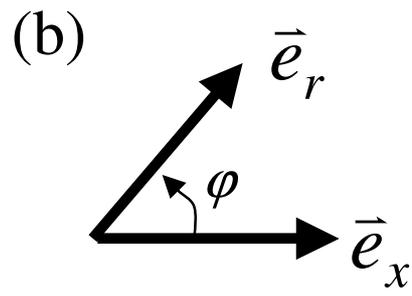
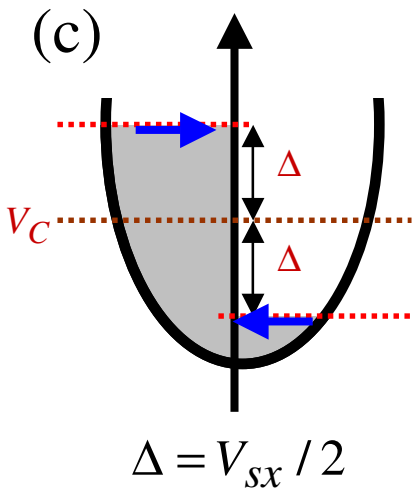
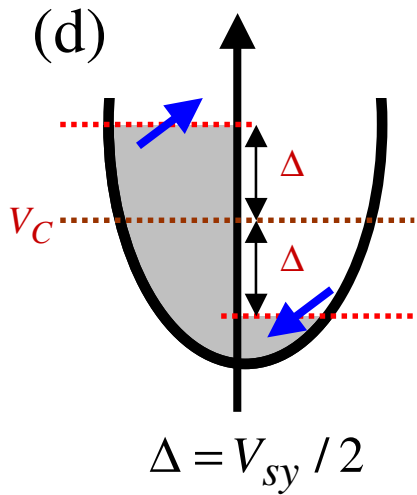
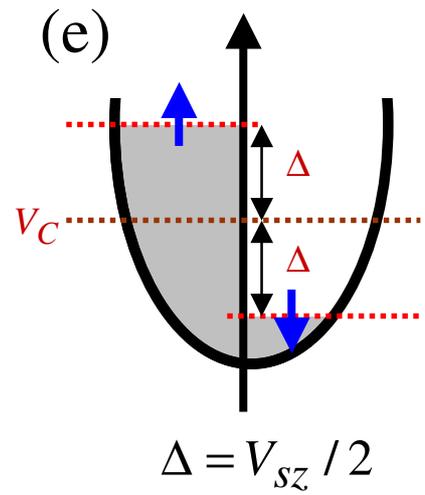
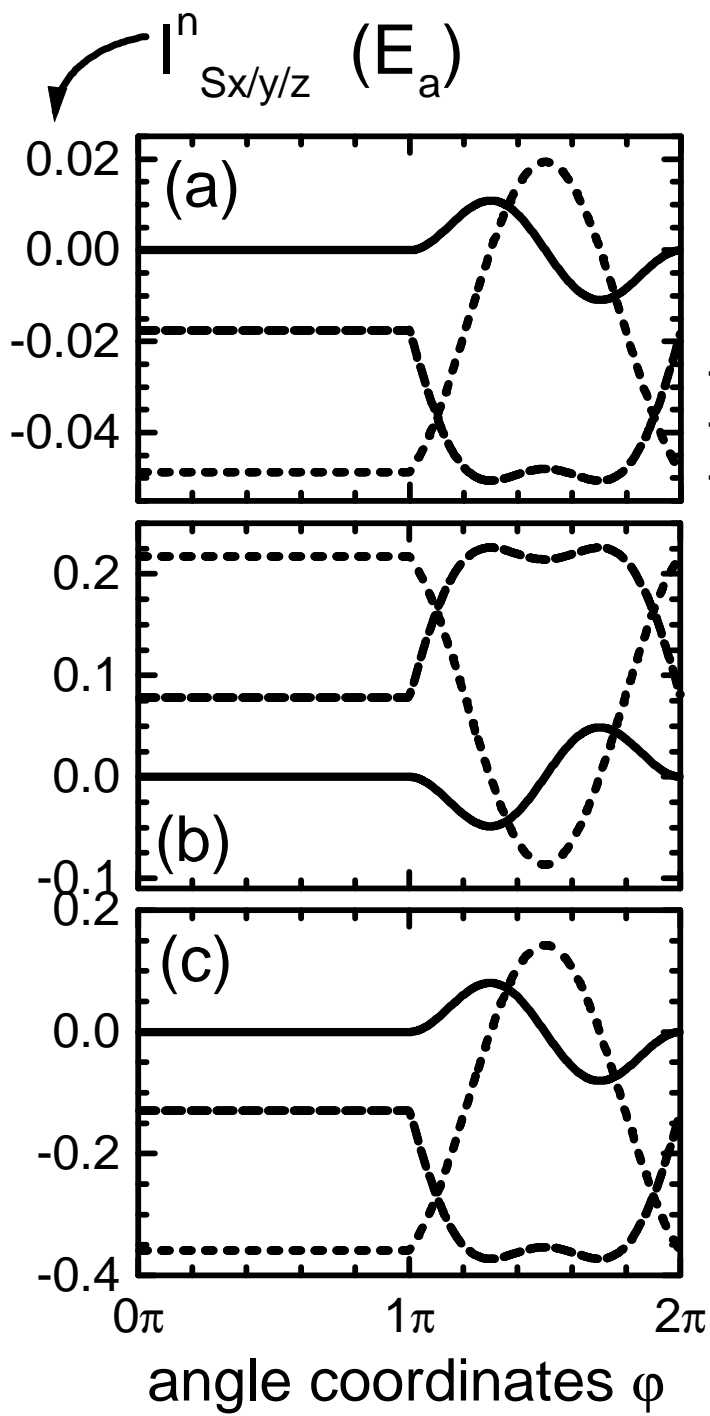
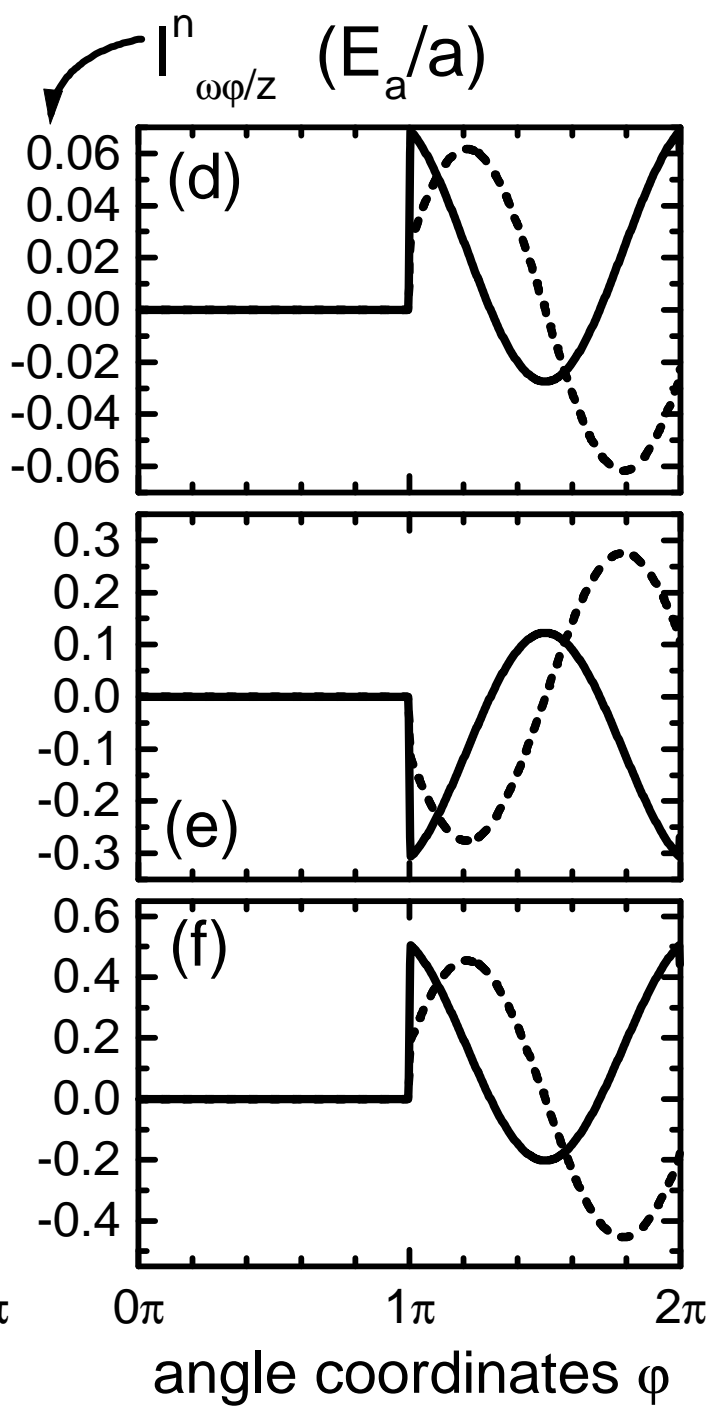
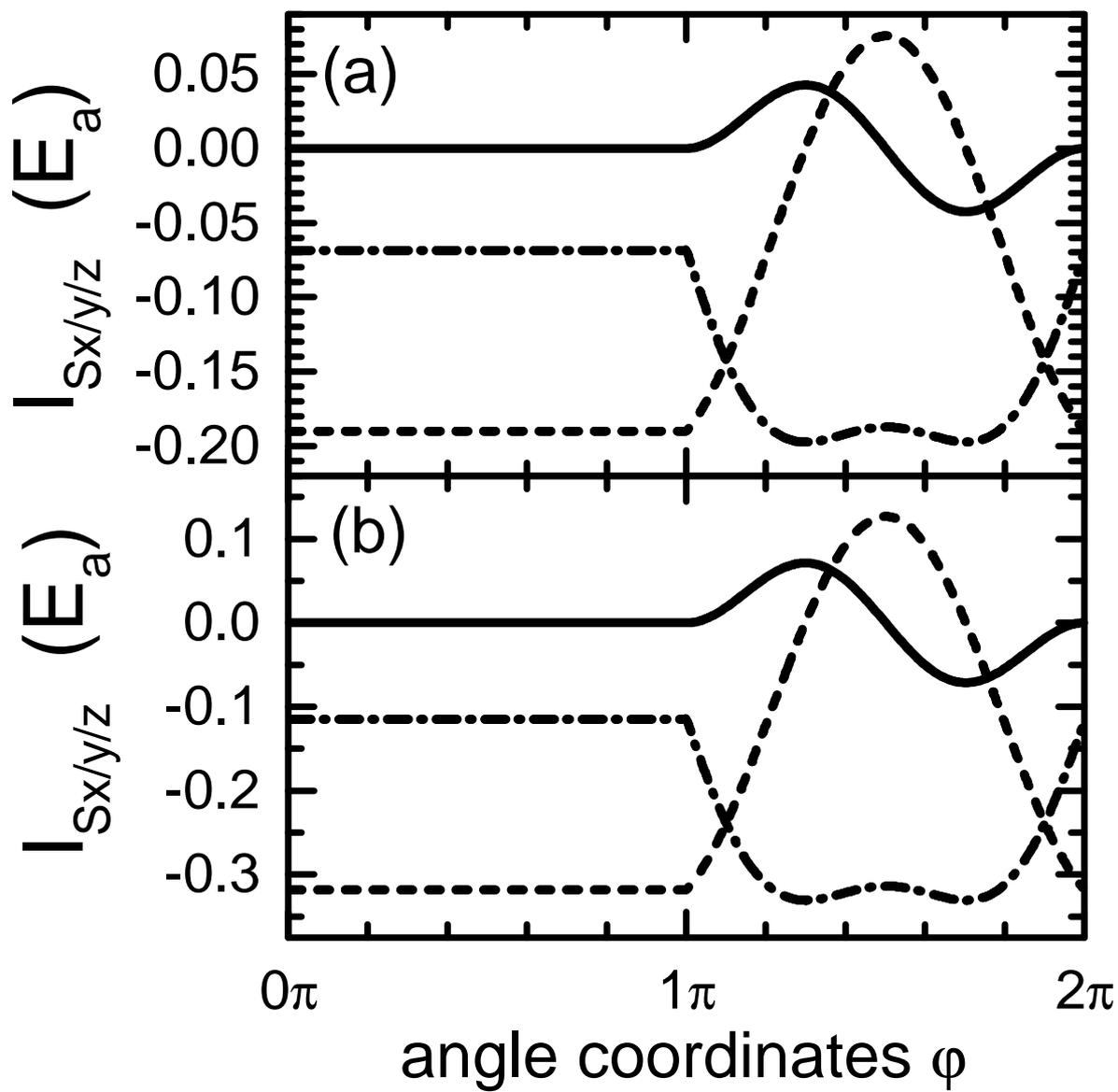
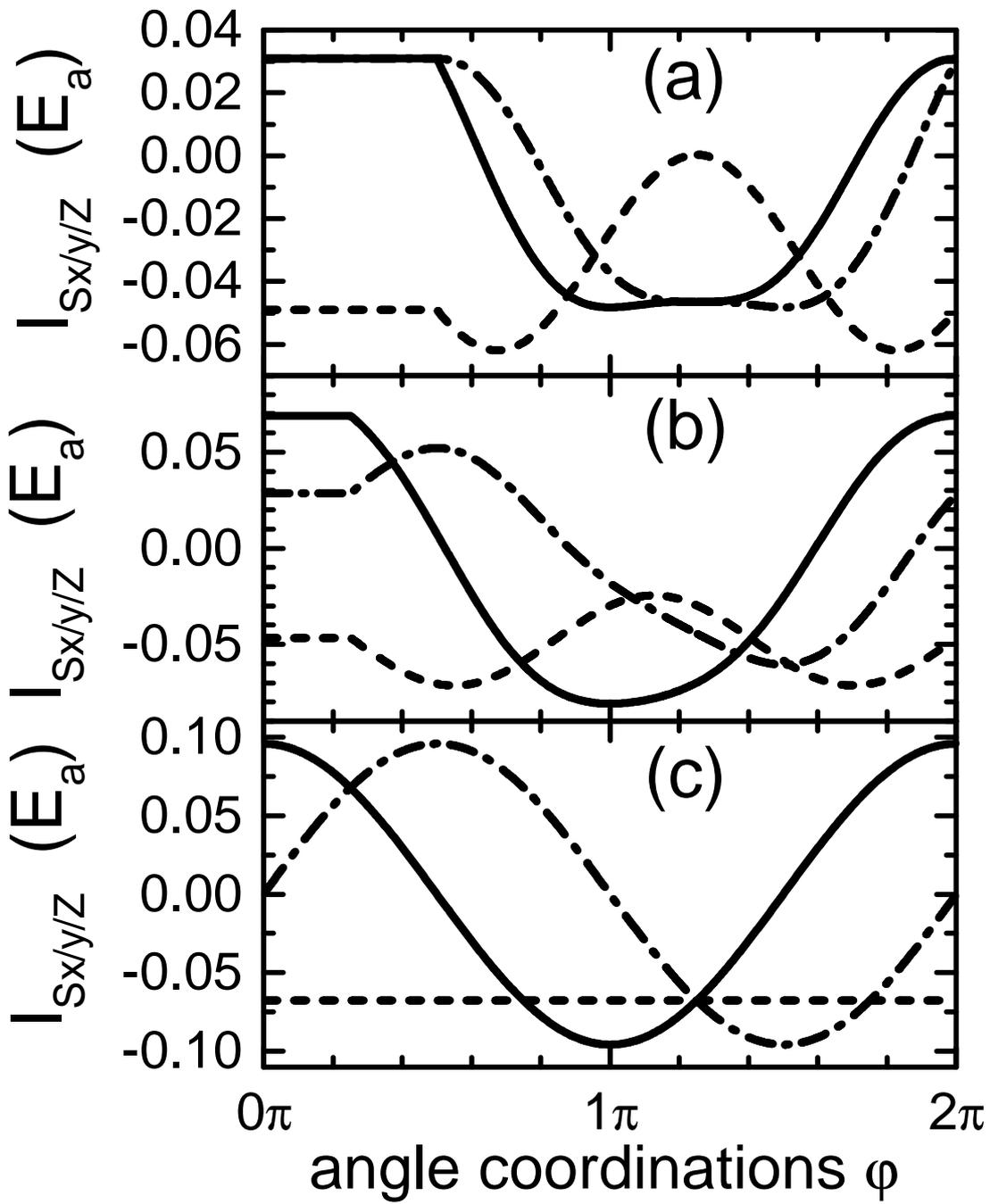
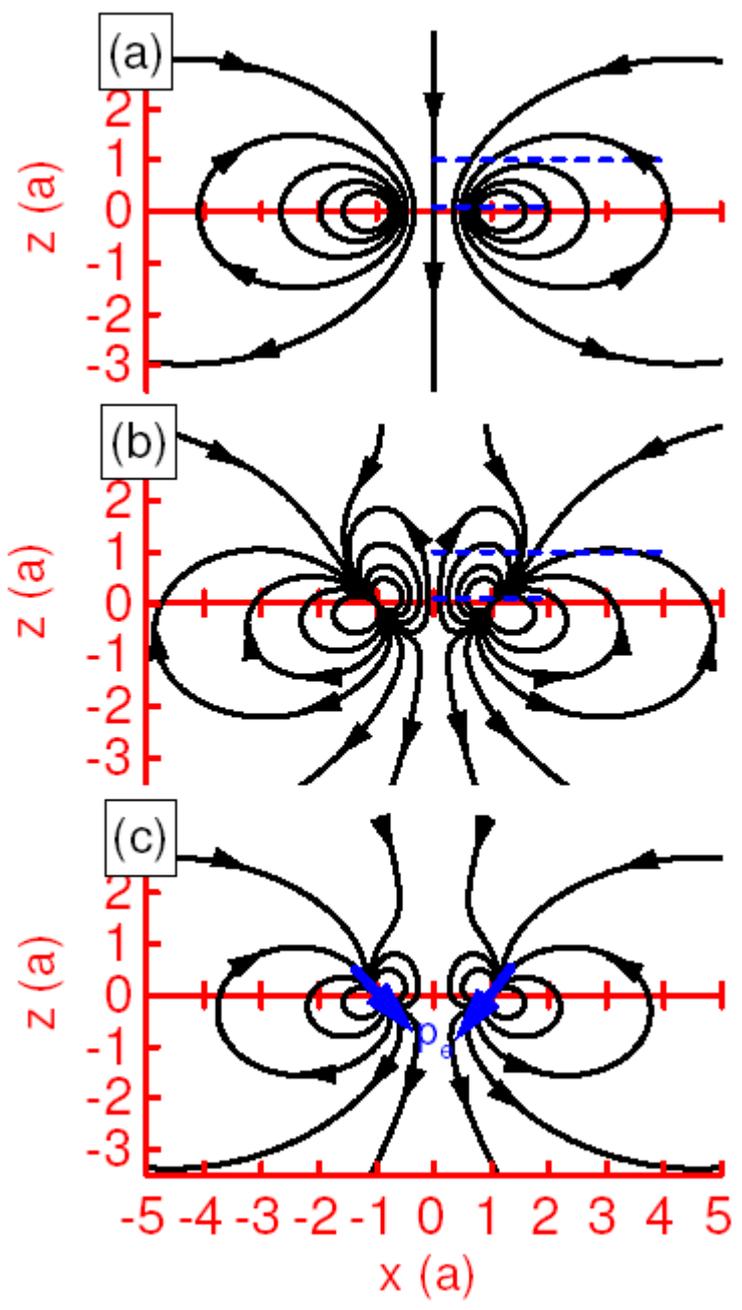
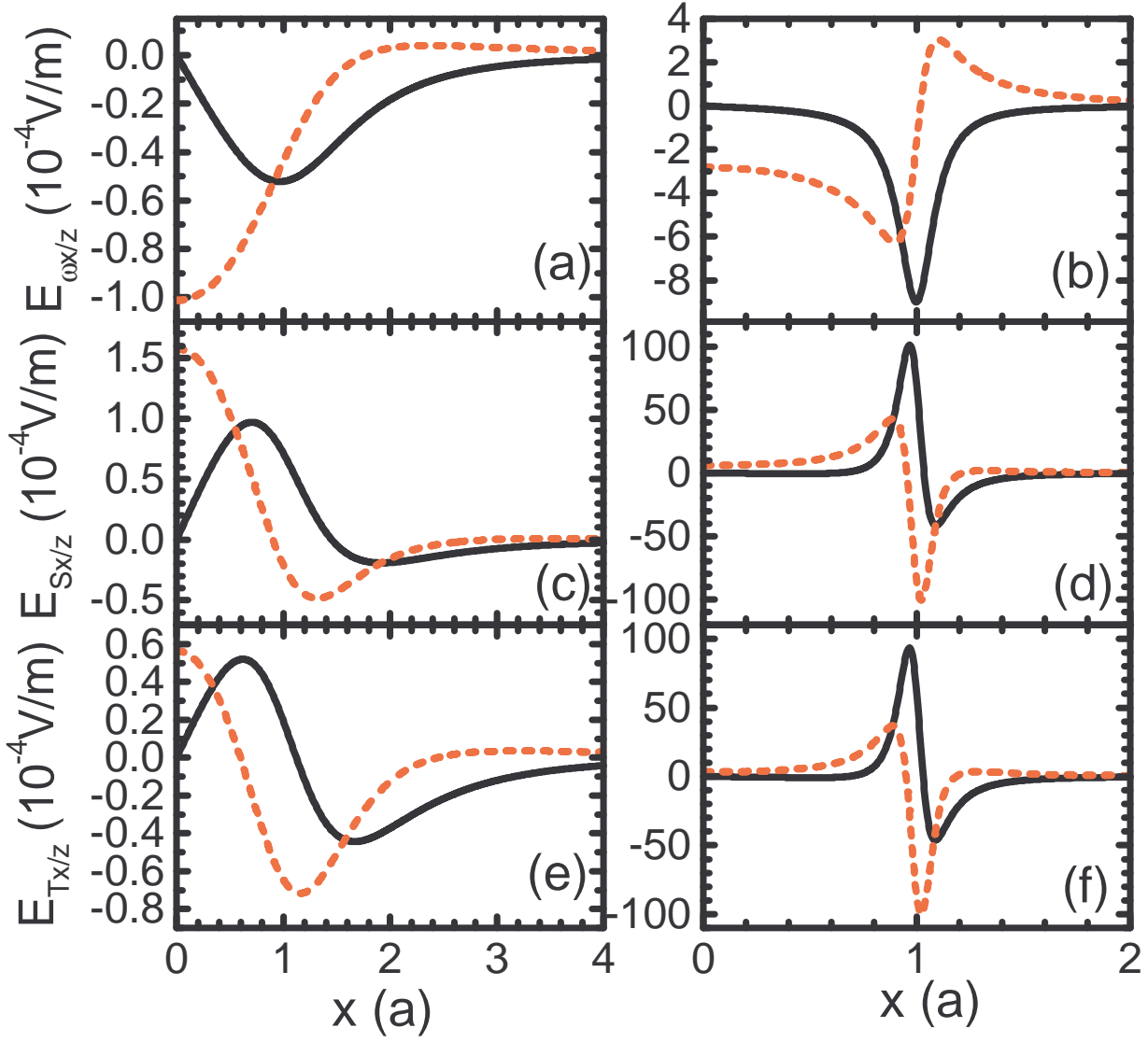
\end{document}